\documentclass[superscriptaddress, nofootinbib, amsmath,amssymb,aps, twocolumn]{revtex4-2}

\usepackage{physics}
\usepackage{soul}
\usepackage{graphicx}
\usepackage{dcolumn}
\usepackage{bm}
\usepackage{hyperref}
\usepackage{standalone}
\usepackage{cleveref}
\usepackage{siunitx}
\usepackage{mathtools}
\usepackage{bm}
\usepackage{esvect}
\usepackage{subcaption}

\begin{document}

\title{Dynamically writing coupled memories using a reinforcement learning agent, meeting physical bounds.}

\author{Th\'eo Jules}
\thanks{These authors contributed equally to this work}
\affiliation{Raymond and Beverly Sackler School of Physics and Astronomy, Tel Aviv University, Ramat Aviv, Tel Aviv, 69978, Israel}

\author{Laura Michel}
\thanks{These authors contributed equally to this work}
\affiliation{Laboratoire de Physique de l'\'Ecole Normale Sup\'erieure, ENS, PSL Research University, CNRS, Sorbonne University, Universit\'e Paris Diderot, Sorbonne Paris Cit\'e, 75005 Paris, France}

\author{Ad\`ele Douin}%
\affiliation{Laboratoire de Physique de l'\'Ecole Normale Sup\'erieure, ENS, PSL Research University, CNRS, Sorbonne University, Universit\'e Paris Diderot, Sorbonne Paris Cit\'e, 75005 Paris, France}

\author{Fr\'ed\'eric Lechenault}
\affiliation{Laboratoire de Physique de l'\'Ecole Normale Sup\'erieure, ENS, PSL Research University, CNRS, Sorbonne University, Universit\'e Paris Diderot, Sorbonne Paris Cit\'e, 75005 Paris, France}

\date{\today}

\begin{abstract}

Traditional memory writing operations proceed one bit at a time, where e.g. an individual magnetic domain is force-flipped by a localized external field.
One way to increase material storage capacity would be to write several bits at a time in the bulk of the material.
However, the manipulation of bits is commonly done through quasi-static operations.
While simple to model, this method is known to reduce memory capacity.
In this paper, we demonstrate how a reinforcement learning agent can exploit the dynamical response of a simple multi-bit mechanical system to restore its memory to full capacity.
To do so, we introduce a model framework consisting of a chain of bi-stable springs, which is manipulated on one end by the external action of the agent.
We show that the agent manages to learn how to reach all available states for three springs, even though some states are not reachable through adiabatic manipulation, and that both the training speed and convergence within physical parameter space are improved using transfer learning techniques.
Interestingly, the agent also points to an optimal design of the system in terms of writing time.
In fact, it appears to learn how to take advantage of the underlying physics: the control time exhibits a non-monotonic dependence on the internal dissipation, reaching a minimum at a cross-over shown to verify a mechanically motivated scaling relation.   
\end{abstract}

\maketitle

\def\thefootnote{}\footnotetext{Contact: theo.jules.physics@gmail.com}\def\thefootnote{\arabic{footnote}}

\section*{Introduction}

At first sight, memory seems like a fragile property of carefully crafted devices. 
However, upon closer inspection, various forms of information retention are present in a wide array of disordered systems~\cite{keim2019memory}.
Governed by a very rugged and complex energy landscape, such systems usually display history-dependent dynamical~\cite{kovacs1963glass, prados2014kovacs, jules2020plasticity} and static~\cite{Matan2002, Diani2009} responses.
Fundamentally, even a single hysteresis cycle can be seen as an embryonic form of memory, as theorized by the Preisach model~\cite{preisach1935magnetische, mayergoyz1986mathematical}.
Combining these cycles into larger structures generates multi-stable systems showing memory capacity, such as the well-known return point memory where the system can remember and return to previously visited states~\cite{Barker1983a, Deutsch2004, mungan2019structure, keim2020global, Keim2021a}.
This basic model for memory has drawn recent interest in various areas of physics including spin ice
~\cite{libal2012hysteresis}, cellular automata~\cite{goicoechea1994hysteresis}, crumpled sheets~\cite{bense2021complex}, glassy ~\cite{lindeman2021multiple}, plastic~\cite{puglisi2002mechanism, regev2021topology} and granular~\cite{keim2020global, Keim2021a} systems, origami bellows~\cite{yasuda2017origami, jules2022delicate}. 
Interestingly, the model's property indicates that information can be written and read from the underlying system, making the internal memory mechanism adequate to store information.
However, the Preisach model is grounded in a quasi-static framework, for adiabatic transformations.
As a result, the specific characteristics of each hysteretic cycle~\cite{terzi2020state} or the addition of internal coupling~\cite{VanHecke2021a, bense2021complex, jules2022delicate} can considerably shrink the set of reachable states and thus the storage capacity of the device. 
In this paper, we show that a controller, in the form of a reinforcement learning agent, can take advantage of the {\it dynamics} to reach all stable states, including adiabatically inaccessible states, effectively restoring the memory of the system to its full capacity.
We base our study on a model framework akin to that introduced in~\cite{puglisi2002rate, jules2022delicate}, i.e., a chain of bi-stable springs with three coupled units and internal dissipation.
After successfully training the agent on a specified set of physical parameters, we demonstrate that transfer learning~\cite{pan2009survey, taylor2009transfer} accelerates the training on different parameters and extends the region of parameter space that leads to learning convergence.
Finally, we investigate the change of the dynamical protocol proposed by the trained control process for a single transition between two states as a function of the dissipation's amplitude.
The transition duration presents a minimum for a critical value of the dissipation that appears to verify a physically motivated scaling relation, pointing to the fact that the agent learns how to harness the physics of the system to its advantage. 

\section*{Model for a chain of bi-stable springs}

\begin{figure}[htb]
    \centering
    \includegraphics[width=1.\linewidth]{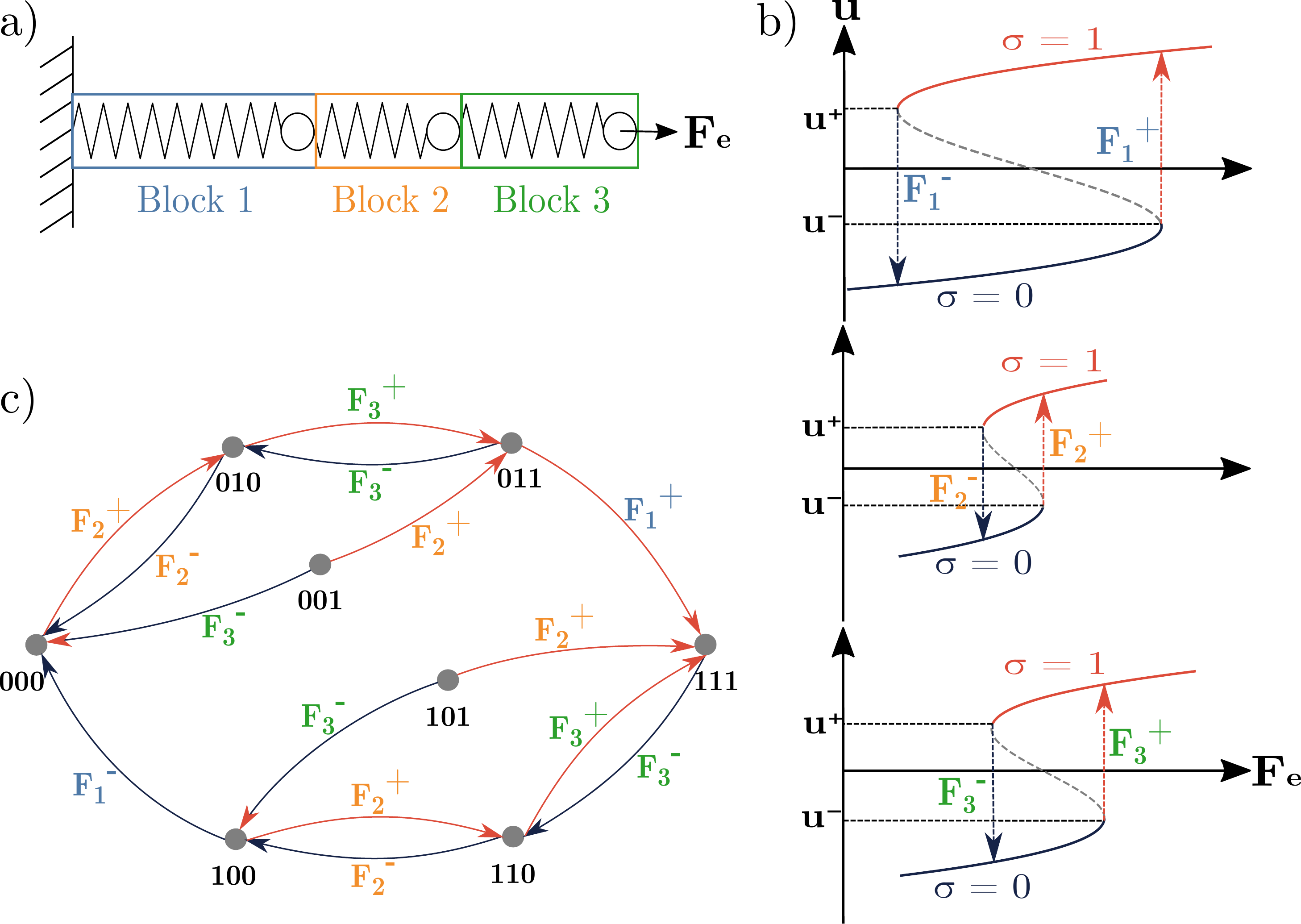}
    \caption{Model for a chain of three coupled bi-stable spring-mass units.
    a) Schematics view of the model. The first unit is attached to a fixed wall and an external force $F_e$ is applied to the last one. 
    b) Deformation $u$ of all three bi-stable springs under external load $F_e$ with the specific choice of disorder $\delta_1^{(0)}$ = 0.050, $\delta_1^{(1)}$ = 0.050, $\delta_2^{(0)}$ = 0.040, $\delta_2^{(1)}$ = 0.020, $\delta_3^{(0)}$ = 0.030, $\delta_3^{(1)}$ = 0.045 (m). The switching fields $F_i^\pm$ are defined through Eq.~[\ref{eq:ThresholdForce}]. 
    c) Transition graph where the nodes represent the stable configurations and the arrows the quasi-statically achievable transitions between them. The states $001$ and $101$ are "Garden of Eden" states (see main text) with this choice of disorder.}
    \label{fig:Model}
\end{figure}

We consider a one-dimensional multi-stable mechanical system composed of $n$ identical masses $m$ connected by bi-stable springs in series, as shown in Fig.~\ref{fig:Model}~a).
For each spring $i$, we set a reference length $l_i$ such that the deformation $u_i$ of the spring reads
\begin{align}
u_i(t) = x_{i}(t)-x_{i-1}(t)-l_i,
\label{eq:DispSpring}
\end{align}
where $x_i(t)$ is the position of the $i$-th mass at time $t$.
The tunable bi-stability of each spring is achieved through a generic quartic potential. We thus obtain the cubic force equations
\begin{align}
F_i(u_i) = -k u_i(u_i+\delta_i^{(0)})(u_i-\delta_i^{(1)}),
\label{eq:ForceSpring}
\end{align}
with $k > 0$ is the stiffness of the spring, $\delta_i^{(0)} > 0$ and $\delta_i^{(1)} > 0$ correspond to meta stable configurations $0$ and $1$, and $u_i = 0$ corresponds to an unstable equilibrium.
Other works on similar systems also considered trilinear forms~\cite{puglisi2002rate, puglisi2002mechanism}.
Still, our choice generates a smooth mechanical response and keeps its amplitude moderate for low deformation.
This behavior plays a crucial role in the scaling analysis that we will present later.
The first spring of the chain is attached to a fixed wall, a condition that imposes $x_0 = 0$, and we apply an external force $F_e$ to the last mass.
Finally, we consider that the system is bathed in an environmental fluid, resulting in a viscous force $F_{f, i}$ being exerted on each mass such that
\begin{align}
F_{f, i} = - \eta \dot{x_i},
\label{eq:Viscou}
\end{align}
where the dot indicates the derivative with respect to time $\dot{x} = \frac{dx}{dt}$ and $\eta$ is a viscous coefficient.

Let us first consider the case $n=1$ where the system is a single bi-stable spring attached to a mass.
The solutions for mechanical equilibrium, schematized in Fig.~\ref{fig:Model}~b), displays two critical amplitudes $F^{\pm}$ such that
\begin{align}
F^{\pm} &= -F(u_{\mp}) \label{eq:ThresholdForce}\\
u_{\pm} &= \frac{1}{3}\left(\delta^{(1)}-\delta^{(0)} \pm \sqrt{(\delta^{(0)})^2 + (\delta^{(1)})^2+\delta^{(0)}\delta^{(1)}}\right) \label{eq:ThresholdDef}
\end{align}
where $u_{\pm}$ are both solutions of $\dv{F}{u}=0$. 
Note that $F^-$ is negative and $F^+$ is positive since we defined both $\delta^{(0)}$ and $\delta^{(1)}$ to be positive. 
These two forces are essential to describe the stability of the system.
Indeed, we observe two branches $\sigma$ of stable configurations, that we call {\tt 0} (for $u < 0$) and {\tt 1} (for $u > 0$).
When $F$ remains in the range $[F^-, F^+]$, both branches present a solution for the mechanical equilibrium $F(u) + F_e = 0$: the system has two stable states.
However, as soon as $F_e$ gets beyond this range, either the branch $\sigma = {\tt 0}$ (for $F_e < F^-$) or $\sigma = {\tt 1}$ (for $F_e > F^+$) vanishes.

These dynamical properties lead to an hysteresis cycle that can be described with a simple experiment.
We start with the spring at rest and $\sigma = 0$, \emph{i.e.} $u=-\delta^{(0)}$, and slowly pull on its free end, increasing $F_e$.
As a response, the system slightly stretches and $u$ increases.
This continues until $F_e = F^+$ where the branch $\sigma = {\tt 0}$ disappears. 
At this point, if we continue to increase $F_e$, the system necessarily jumps to the branch $\sigma = {\tt 1}$, with $u > 0$, in order to maintain mechanical equilibrium.
If we now decrease $F_e$, $u$ decreases accordingly until $F_e = F^-$ where the system has to jump back to the original branch $\sigma = {\tt 0}$.
If we only consider the stability branches, the described hysteresis cycle corresponds to a so-called hysteron, the basic block of the Preisach model~\cite{Preisach1935}.
In a Preisach model, a set of $n$ independent hysterons is actuated through an external field.
Each hysteron has two states, {\tt 0} and {\tt 1}, and two switching fields, $F_i^+$ and $F_i^-$, that characterize when each hysteron $\sigma_i$ switches between states.
As long as the switching fields are unique and the hysterons are independent, \emph{i.e.} the switching fields do not change with the state of the system, and the Preisach model is able to predict the possible transitions between the $2^n$ configurations.
Our system, $n$ bi-stable spring/mass in series, under quasi-static actuation, fulfills the local mechanical equilibrium and independence assumptions.

To get a better visual representation of the model's predictions, the  quasi-statically achievable transitions between states are modeled as the edges of a directed graph, where the nodes represent the stable configurations.
The allowed transitions correspond to the switch of a single hysteron.
The exact topology of the transition diagram depends on the relative values of the switching fields~\cite{terzi2020state}.
If the order for $|F_i^-|$ and $|F_i^+|$ is the same for all the coupled hysterons, the system can reach every configuration from any other one.
Otherwise, there exists isolated configurations which can never be reached again once left.
As is customary, we call these unreachable configurations "Garden of Eden" (GoE) states (see~\cite{jules2022delicate} for a historical account).
An illustration for the case $n=3$ is given in Fig.~\ref{fig:Model}~c) where both ${\tt 001}$ and ${\tt 101}$ are GoE states.
The existence of GoE states severely limits the total number of reachable states, and as a result, the amount of information that can be stored in the system.

A solution to overcome this limitation is to break the quasi-static assumption and to take advantage of the {\it dynamics} as a means to reach GoE configurations.
However, this approach foregoes a major upside of the Preisach model, its simplicity.
Indeed, applying Newton's second law of motion to each mass yields a description of the dynamical evolution for the system through a system of $n$ coupled non-linear differential equations
\begin{align}
	\begin{cases}
		m\ddot{x_i} = F_{i}(u_{i}) - F_{i+1}(u_{i+1}) + F_{f, i} & \text{for $0 < i < n$},\\
		m\ddot{x_n} = F_{n}(u_{n}) + F_e(t) + F_{f, n}.
	\end{cases}
\label{eq:DynamicEq}
\end{align}
While providing a clear experimental protocol to make the system change configuration is straightforward in the quasi-static regime, the non-linear response of the springs and the coupling between the equations make a similar analysis a hefty challenge in the dynamical case.
In the following, we demonstrate how the use of artificial neural networks and reinforcement learning unlocks this feat.

\section*{Reinforcement learning to control the dynamic}

Reinforcement Learning (RL) is a computational paradigm that consists in optimizing, through trial-and-error, the actions of an agent that interacts with an environment.
RL has been shown to be an effective method to control multistable systems with nonlinear dynamics~\cite{gadaleta1999optimal, gadaleta2001learning, wang2021constrained, pisarchik2014control}.
In RL, the optimization aims to maximize the cumulative reward associated with the accomplishment of a given task.
At each step $t$, the environment is described by an observable state $s_t$.
The agent uses this information, in combination with a policy $\mu$, to decide the action $a_t$ to be taken, \emph{i.e.} $\mu(s_t) = a_t$.
This action brings the environment to a new state $s_{t+1}$, and grants the agent with a reward $r_t$ quantifying its success with respect to the final objective. An {\it episode} ends when that goal is reached or, if not, after a finite time.
The goal of training is to learn a policy that maximizes the agent's cumulative reward over an episode.
When the control space is continuous, one can resort to an {\it actor-critic} architecture~\cite{konda1999actor}, which is based on two Artificial Neural Networks (ANN) learning in tandem.
One network, called the actor, generates a sensory-motor representation of the problem in the form of a mapping of its parameter space $\theta$ into the space of policies, such that $\mu = \mu_{\theta}$. This operation makes the exploration of the continuous spectrum of choices computationally tractable.
The optimization of the actor necessitates an estimation of the expected reward at long time.
This is the role of the second ANN, the critic, which learns to evaluate the decisions of the actor, and how it should adjust them. 
This is done though the same bootstrapping of the Bellman's equation as that used in Q-learning~\cite{grondman2011efficient}.
The successive trials - resulting in multiple episodes - are stacked in a finite memory queue (FIFO), or replay buffer, and after each trial the ANN tandem is trained on that buffer, thus progressively improving their decision policy and the quality of the memory.
The precise architecture of the algorithm we used is based on Twin Delayed Deep Deterministic Policy Gradient~\cite{TD3} and is detailed in the {\it Materials and Methods} section.

In our system, the environment consists in the positions and velocities of the masses, an action is a choice for the values of the force applied to the last mass in the chain at time $t$, and the goal is to bring the system close to a given meta-stable memory state in a given time $t_{\max}$ - close enough that it cannot switch states if let free to evolve. 
At the start of an episode, the environment is randomly initialized, and a random target state is set.
The information provided as an input to the networks includes the position $x$ and velocity $\dot{x}$ of all the masses in addition to the one-hot encoded target configuration.
Then the policy decides on the force $F_e$ applied to reach the next step.
We reduce the space of possible actions by limiting the amplitude of the force to be lower or equal to one newton ($|F_e| < 1 N$).
The evolution of the system is simulated by solving the differential equations~(\ref{eq:DynamicEq}) numerically with a Runge-Kutta method of order 4.
The reward from this action is computed relative to the newly reached state: we give a penalty ($r_{t} < 0 $) with an amplitude proportional to the velocity of the masses and to the distance of the masses from their target rest positions.
After every step, the replay buffer is updated with the corresponding data.
The critic is optimized with a batch of data every step, while the actor is optimized every two steps.
The episode stops if the system is sufficiently close to rest in the correct configuration, in which case a large positive reward ($r_{t} \gg 1$) is granted, or after $t_{\max}$.
Then a new episode is started, and the algorithm is repeated for a predefined number of episodes.
More details are available in the {\it Materials and Methods}.

\begin{figure}[h!]
    \centering
    \includegraphics[width=0.9\linewidth]{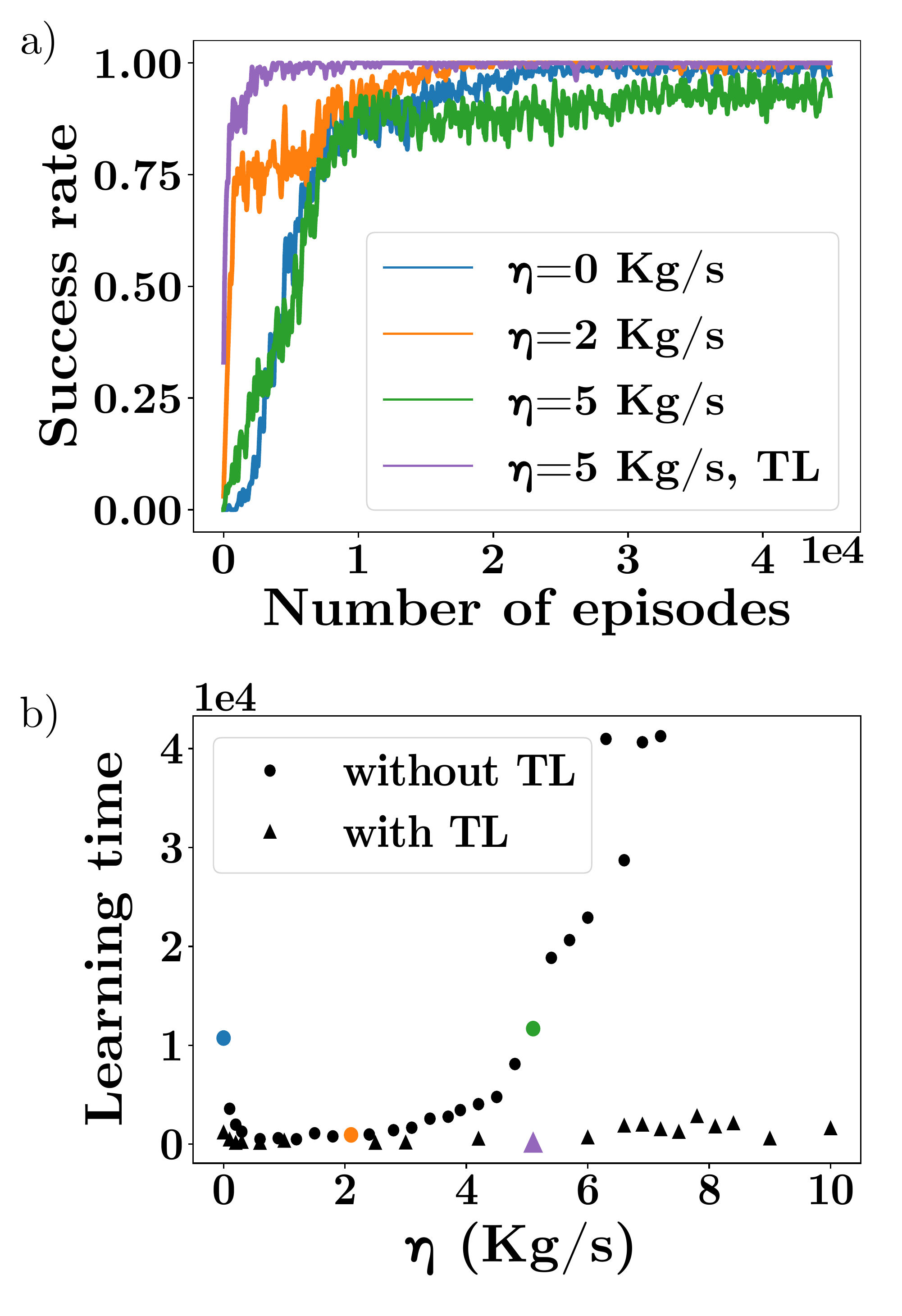}
    \caption{Training dynamics of the RL agent on the model presented in Fig.~\ref{fig:Model} and different values of the viscous coefficient.
    a) Evolution of the success rate during training. The success rate is defined as the number of times the agent succeeds in reaching the target configuration over the last 100 episodes. 
    For the blue, green, and orange curves, the ANN was initialized randomly. 
    For the purple curve, the ANN was initialized using the weights of a previously learned model with $\eta$=4 Kg/s. 
    b) Learning time with respect to the viscous coefficient with and without Transfer Learning. The learning time is defined as the number of episodes it takes for the success rate to reach the threshold value 0.8 for the first time.}
    \label{fig:TransferLearning}
\end{figure}

Using the described method, we trained our ANNs on a chain of three bi-stable springs specifically designed to display GoE states, as detailed in the {\it Materials and Methods} section and illustrated in Fig.~\ref{fig:Model}~c.
Interestingly, the networks achieve a 100\% success rate at reaching any target state - including the GoE states - in less than 10 000 episodes, as shown in Fig.~\ref{fig:TransferLearning}.~a). 
We thus accomplished our initial objective and designed a reliable method that produces protocols for the transitions to any stable configurations, restoring the memory of the device to its full capacity.

In order to evaluate the robustness of the decision making process of the policy and gain insights on the mechanisms involved, we study how the physical parameters of the system influence the designed solutions.   
We chose to focus on a quantity that deeply affects the dynamics of the masses and has simple qualitative interpretation: the viscous coefficient $\eta$.
To observe its effect on the designed policy, we trained agents with random initialization of weights for a range of $\eta$ while keeping all other physical parameters consistent.
The learning time, defined as the number of episodes before the success rate reaches 80$\%$ during training, is shown in Fig.~\ref{fig:TransferLearning} b) as a function of $\eta$.
Even though the algorithm manages to learn the transitions for a wide range of $\eta$, the learning time varies significantly.
Notably, the learning time gets longer for very low viscous coefficients (here $\eta < 0.1$ kg/s) but also seems to diverge at very high $\eta$.

Due to the continuous nature of the system, we expect small modification of the physical parameters not to catastrophically change the dynamics of the system.
With this assumption, we employed Transfer Learning (TL) techniques~\cite{pan2009survey, taylor2009transfer} between runs at different $\eta$ to accelerate the learning phase.
In TL, the ANNs are not initialized with random weights, but with the weights of ANNs already trained on a similar physical system but with a slightly different $\eta$.
The expectation is that some physical principles learned by the algorithm remain applicable for solving the new problem.
We thus slowly increase $\eta$ from from 2 kg/s up to 10 kg/s and decrease it from 2 kg/s to 0 kg/s, applying TL at each increment. 
We observe in Fig.~\ref{fig:TransferLearning}. b) that TL is very effective, dividing by up to 30 the learning time for very high viscosity, and allowing to reach otherwise non-converging regions.
This acceleration of training also allows for a finer discretization of the viscous coefficient exploration while keeping computation time reasonable.

By mixing RL and TL, we generated an algorithm that quickly produces precise transition protocols to any stable state for chains of bi-stable springs, including GoE states.
In the next section, we analyze the properties of the force signals produced by the ANN and investigate how they relate to the dynamics of the system as the viscous coefficient $\eta$ is varied. 

\section*{How damping affects the control strategy}
\label{sec:damping}

\begin{figure}[h]
\centering
\begin{subfigure}{0.5\columnwidth}
		\centering
        \includegraphics[width=1\linewidth]{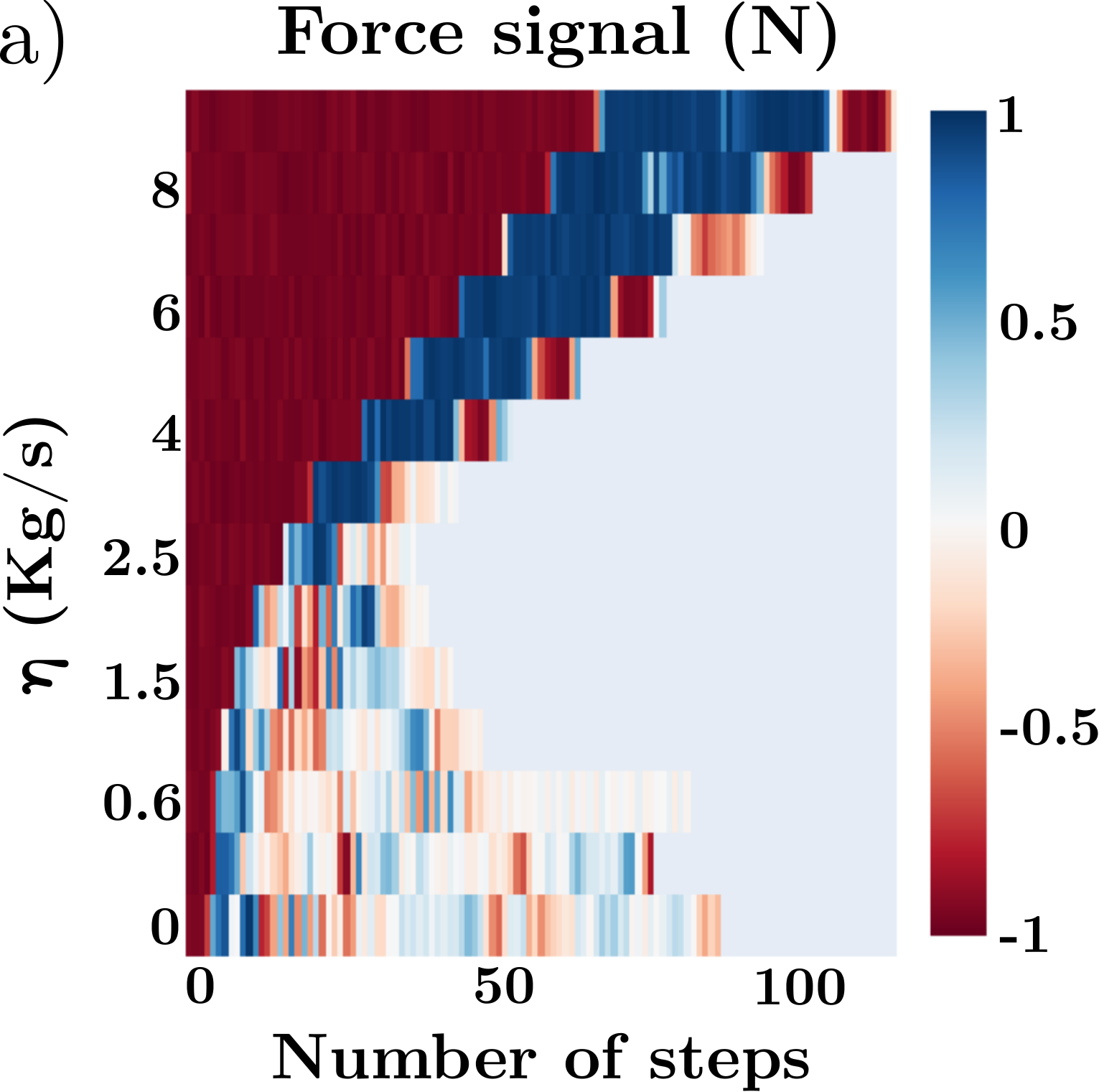}
\end{subfigure}%
\begin{subfigure}{0.5\columnwidth}
		\centering
        \includegraphics[width=1\linewidth]{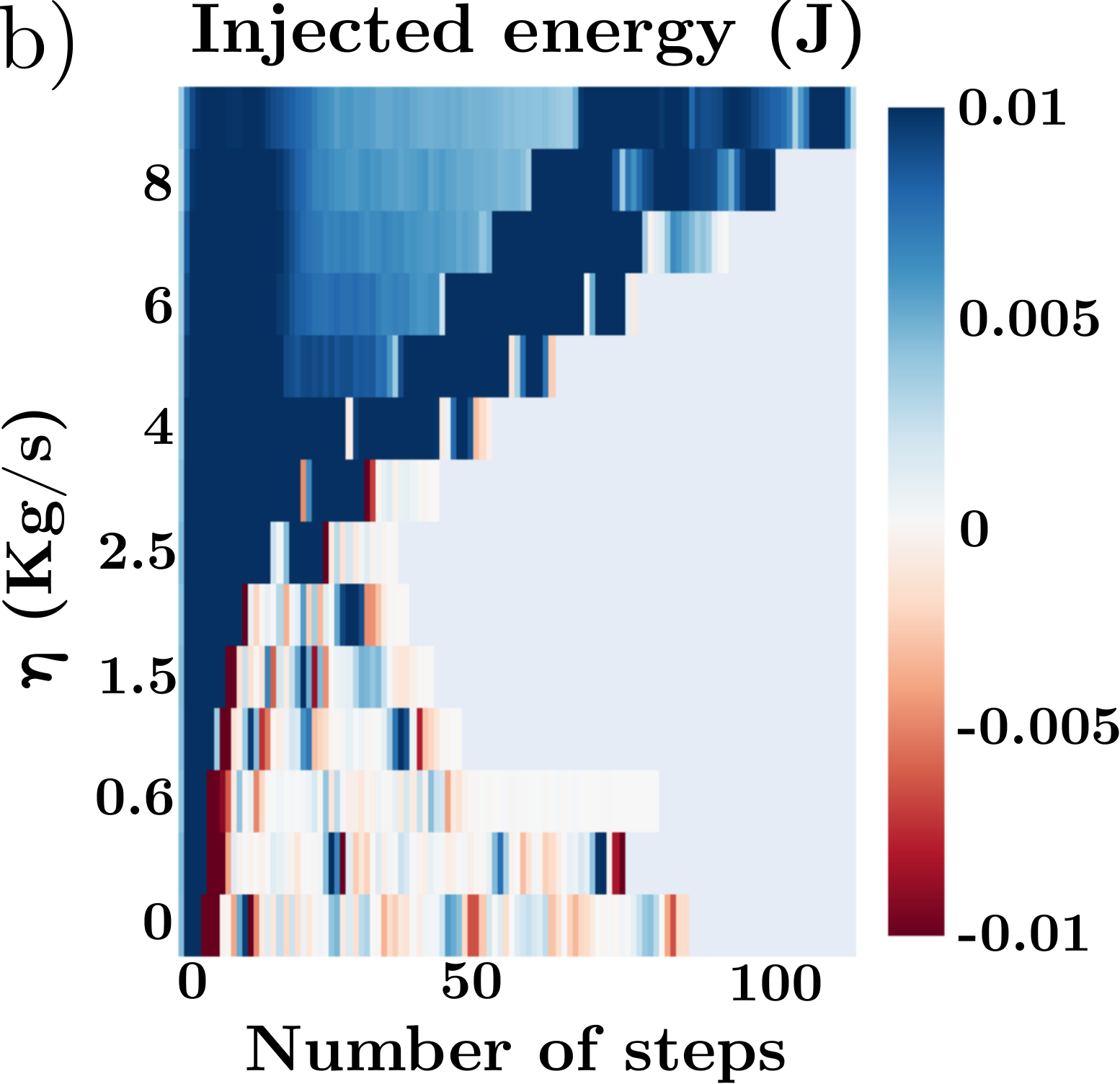}
\end{subfigure}%
        \hfill%
\begin{subfigure}{0.5\columnwidth}
		\centering
        \includegraphics[width=1\linewidth]{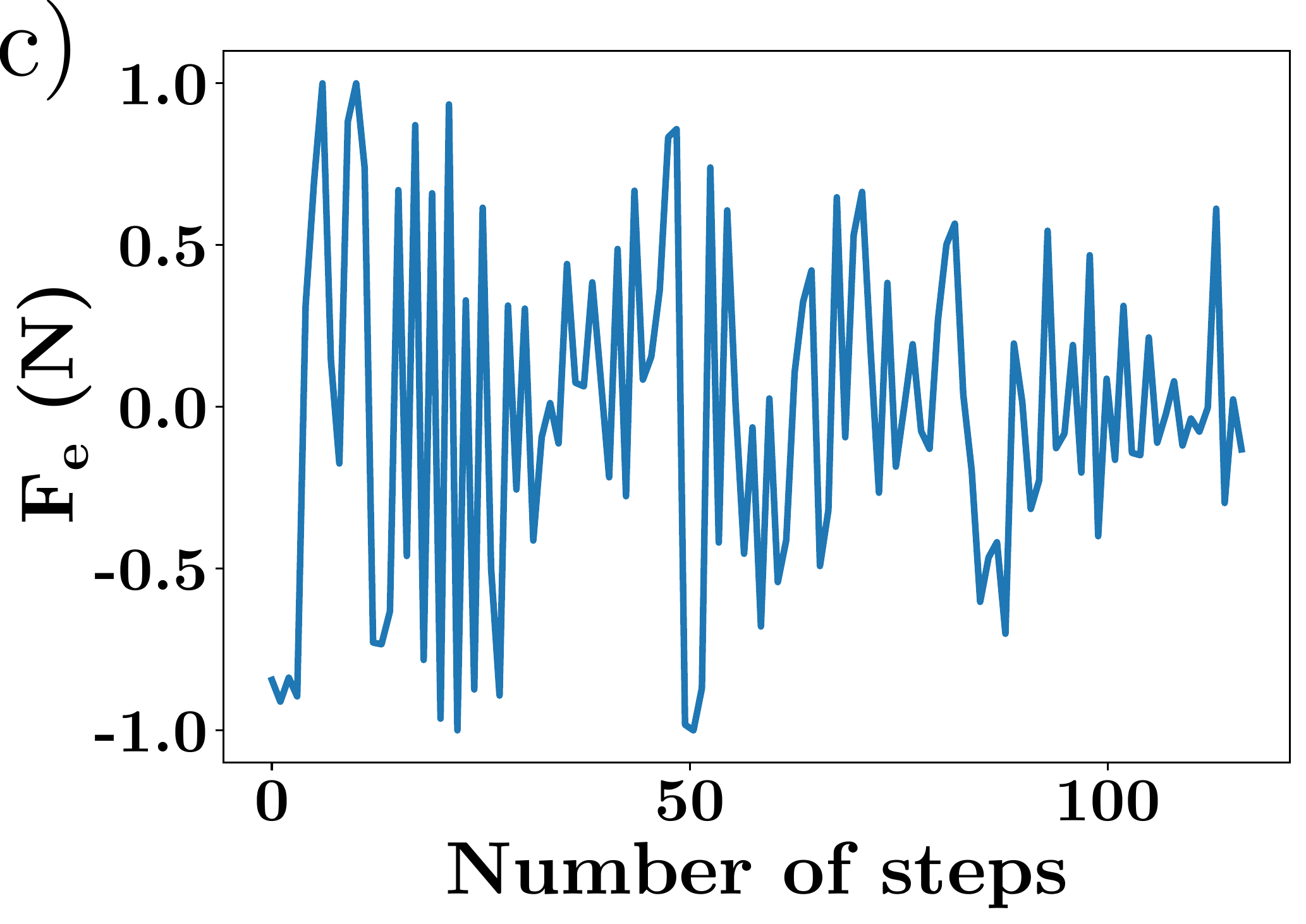}
\end{subfigure}%
    \hfill%
\begin{subfigure}{0.5\columnwidth}
		\centering
        \includegraphics[width=1\linewidth]{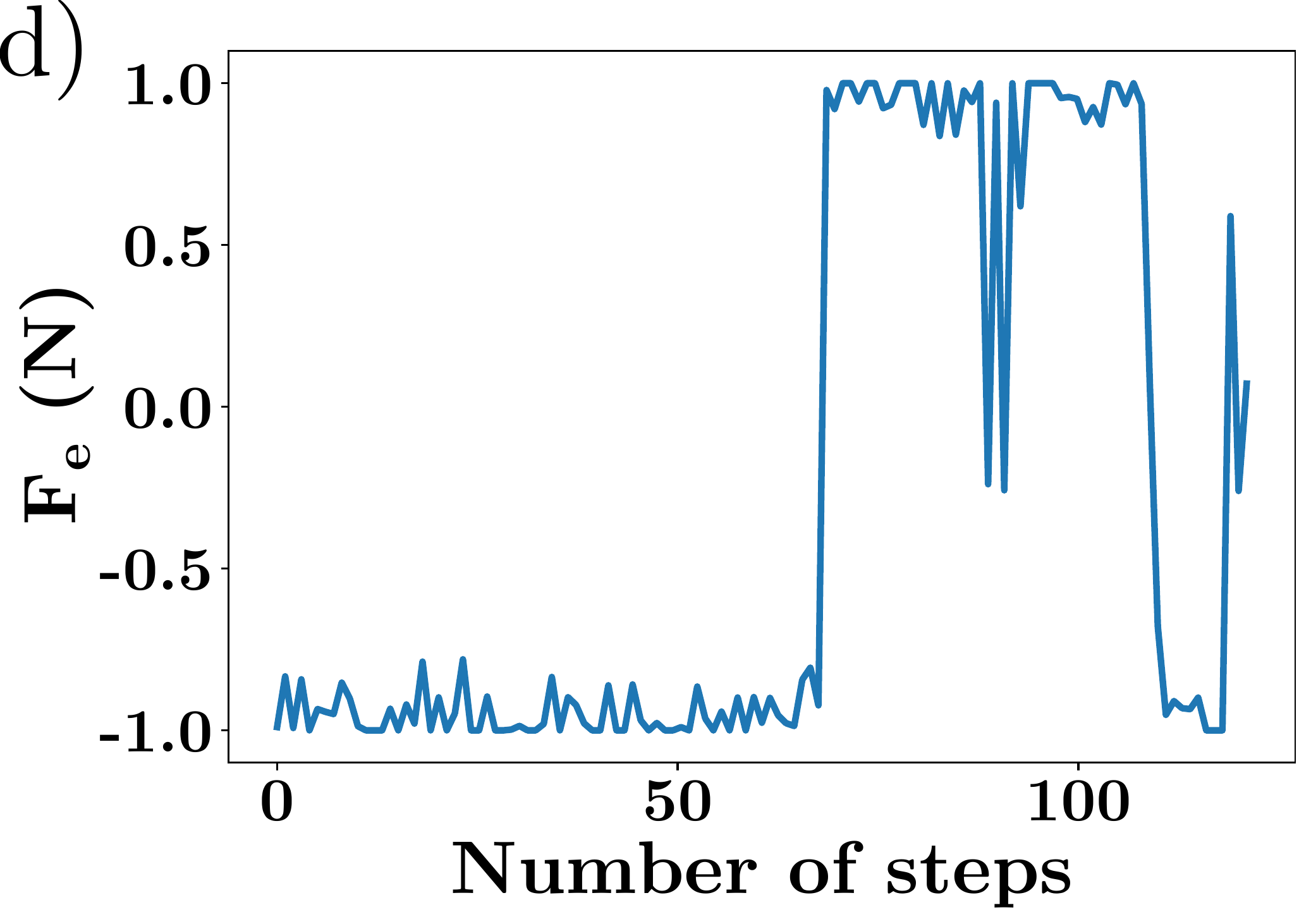}
\end{subfigure}%
    \hfill%
\begin{subfigure}{0.5\columnwidth}
		\centering
        \includegraphics[width=1\linewidth]{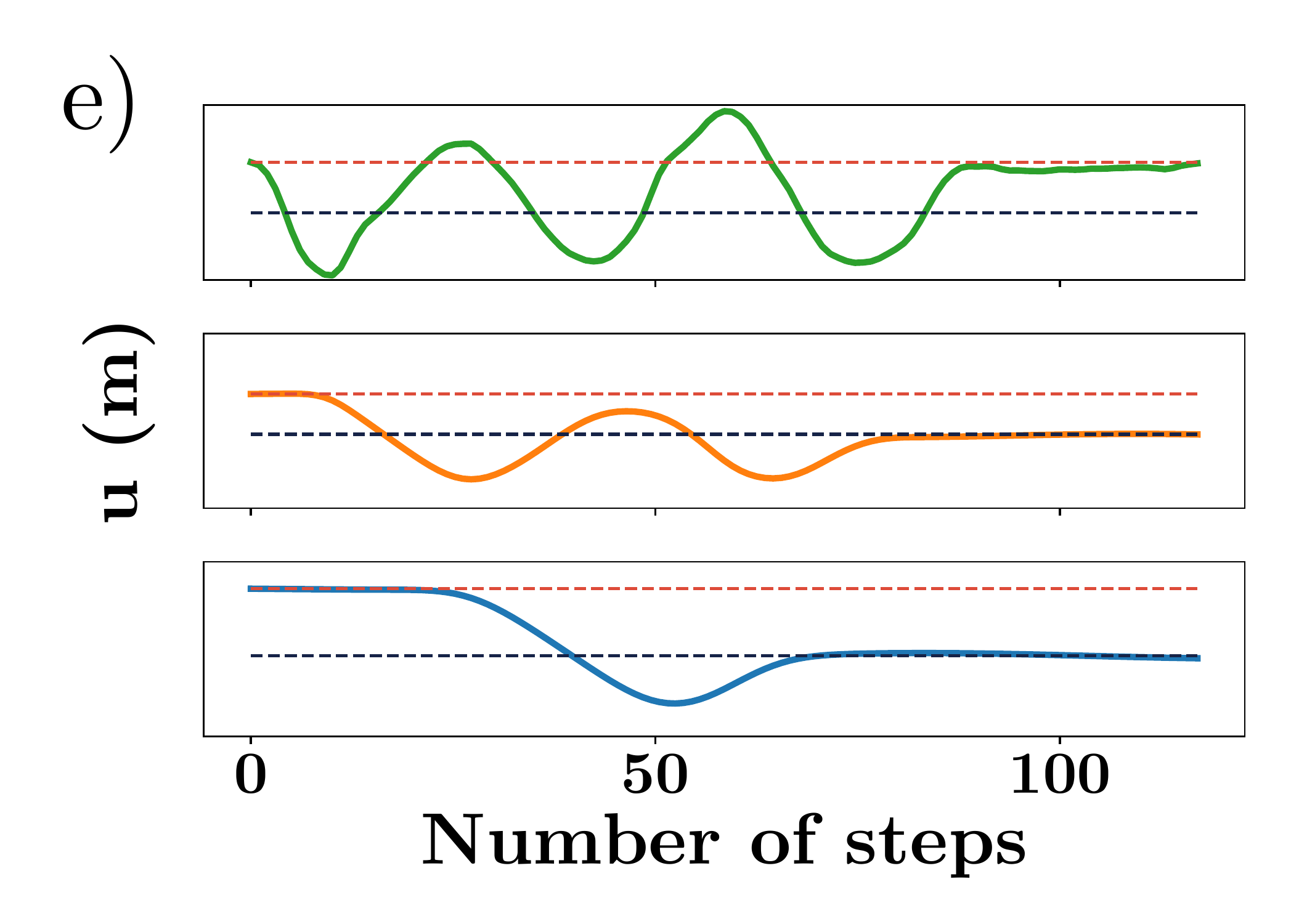}
\end{subfigure}%
    \hfill%
\begin{subfigure}{0.5\columnwidth}
		\centering
        \includegraphics[width=1\linewidth]{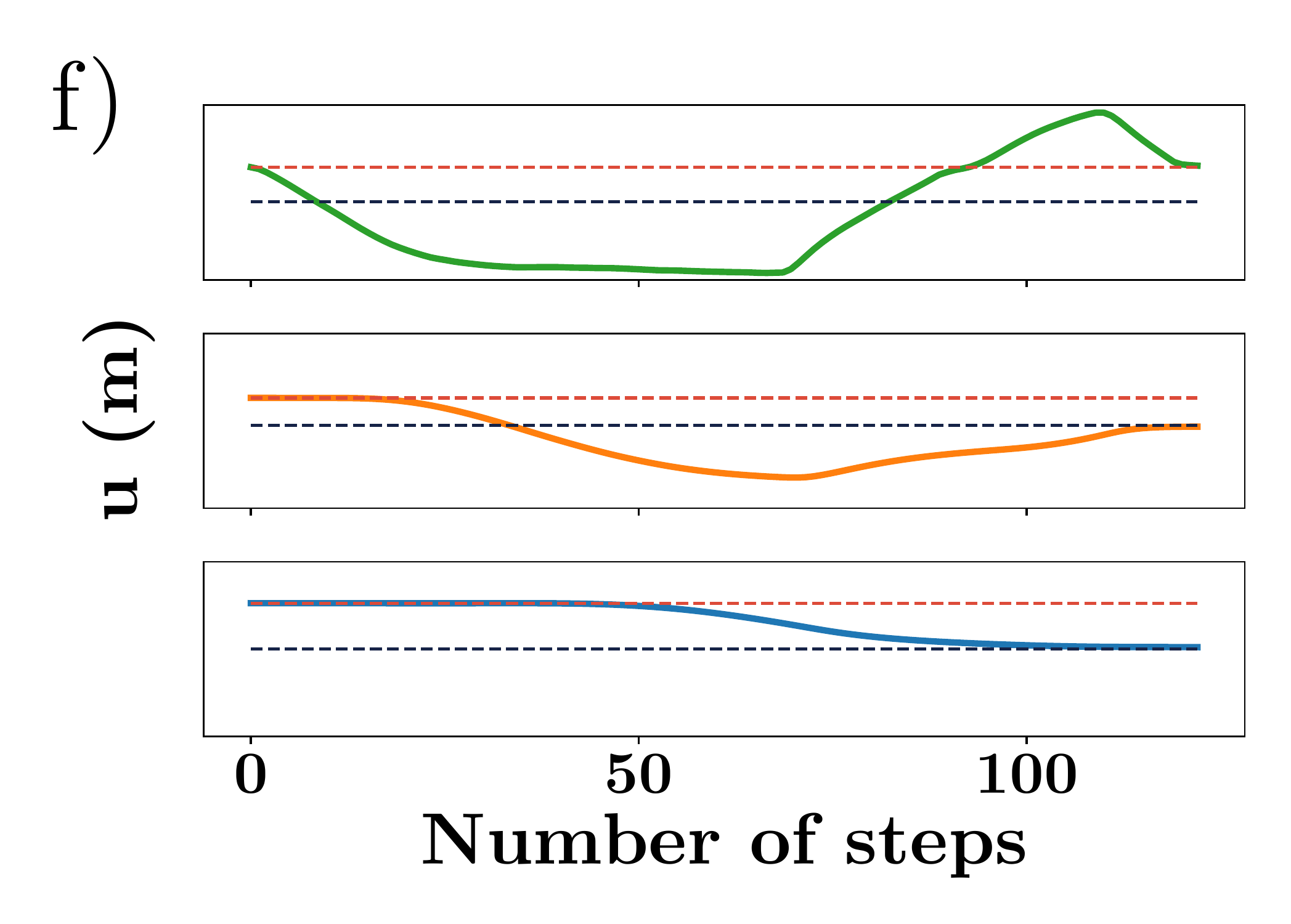}
\end{subfigure}%
    \caption{Analysis of the protocol proposed by the RL agent for the transition ${\tt 111 \rightarrow 001}$ with different values of the viscous coefficient.
    a) Force signal and b) injected energy during the transition. 
    The injected energy is computed by multiplying the chain's elongation with the value of the external force.
    c) Force signal and e) deformation of each bi-stable spring during the transition for $\eta$ = 0 kg/s (low-viscosity regime). 
    The colors blue, orange, and green correspond to the deformation of the first, second, and third spring, respectively.
    The subplots for the deformations all have the same height with edge values [-0.07, 0.19] meter.
    The dashed lines represent the stable equilibria $\delta^{(0)}$ (black) and $\delta^{(1)}$ (red).
    d) Force signal and f) deformation of each bi-stable spring during the transition for $\eta$ = 9 kg/s (high-viscosity regime). 
    The subplots for the deformations all have the same height with edge values [-0.14, 0.24] meter.}
    \label{fig:effect_damping}
\end{figure}

The intensity of the damping impacts the dynamical response of the system, which significantly affects the actuation protocol proposed by the ANN.
To study these variations, we selected a unique transition ({\tt 111 $\rightarrow$ 001}), recorded the signal of the force generated by the agent, and computed the corresponding mechanical energy injected into the system at each time step for different values of $\eta$, as shown in Fig.~\ref{fig:effect_damping}.
Please note that the state {\tt 001} is a GoE state.
We observe that the relation between $\eta$ and the time it takes to reach the target state is non-monotonic.
We identify the minimum episode duration, corresponding to the most efficient actuation, with a critical viscous coefficient $\eta = \eta_c$.
Interestingly, this minimum also marks the transition between two qualitatively different behaviors of the control force: a high-viscosity regime ($\eta > \eta_c$) and a low-viscosity regime ($\eta < \eta_c$) as shown in Fig.~\ref{fig:effect_damping} c) and d).
In the high-viscosity regime, $F_e$ always saturates its limit value, and only changes sign a few times per episode.
Remarkably, each sign change occurs roughly when a mass is placed at the correct position.
In contrast, the force signal in the low-viscosity regime appears less structured, with large fluctuations between consecutive steps.

In order to qualitatively explain these different behaviors, we focus our analysis on the energy transfer between the external controller and the system.
The starting and the final states are stable configurations at rest.
Consequently, they both correspond to local energy minima and the agent has to provide mechanical energy to the system in an effort to overcome the energy barriers between these configurations.
After crossing the barriers, the surplus of kinetic energy has to be removed to slow down the masses and trap them in the well associated with the targeted minimum.
In the low-viscosity regime, the internal energy dissipated due to viscosity is small.
As a result, the protocols require phases where the agent is actively draining energy from the system.
After a short initial phase of a few steps, where much energy is introduced into the system by setting the external load to its maximal value, a substantial fraction of the remainder of the episode involves careful adjustments to remove the kinetic energy.
In the high-viscosity regime on the other hand, the viscosity is able to dissipate the extra energy without further intervention. As it increases, the dynamics slow down, which translates into an increasing episode duration.

While we established the characteristics of the designed protocols in both regimes, we have yet to define a quantitative estimation of the crossover between regimes.

Drawing inspiration from these protocols, we consider a simplified situation where the external force saturates the constraint $F_e(t) = F_{\max}$, and where the switching fields are very small, i.e. $F_{\text{max}} \gg |F_i^{\pm}|,\,i=1, 2, 3$.

Since all the masses start in a stable equilibrium, the mechanical response of the chain to the external load is very soft, at least for small enough displacements at the start of the episode.

Thus, we can approximates the dynamics of the last mass by the ordinary differential equation
\begin{align}
	\tau \ddot{x}_3 \approx v_{\max} - \dot{x}_3,
\label{eq:short_time}
\end{align}
which solves into
\begin{align}
	\dot{x}_3(t) = v_{\max}(1-e^{-t/\tau}).
	\label{eq:sol_short_time}
\end{align}
with a relaxation time $\tau = \frac{m}{\eta}$ and a saturation velocity $v_{\max} = \frac{F_{\max}}{\eta}$.
The relaxation time $\tau$ corresponds to the time it takes for dissipation to take over inertia.
This transition is also associated to a length scale $L_{\eta}$ such that
\begin{align}
	L_{\eta} = \tau v_{\max} = \frac{m F_{\max}}{\eta^2}
    \label{eq:length_scale_viscosity}
\end{align}

Let us now define more precisely what we mean by small displacement. With our assumptions, and due to the asymptotic shape of the potential, the typical relative distance $L_e$ for which the mechanical response becomes of the same magnitude as the external load verifies
\begin{align}
	k L_e^3 \sim  F_{\max}
    \label{eq:validity_inertia}
\end{align}

It is thus clear that if $L_{\eta}<<L_e$, the system will be dominated by dissipation and converge to equilibrium without further oscillations. On the other hand, if $L_{\eta}>>L_e$, neighboring masses will rapidly feel differential forces and inertia will dominate. Interestingly, equating these two length scales allows to point to a critical dissipation at the frontier of these two regimes
\begin{align}
	\eta_c \sim m^{1/2} k^{1/6} F_{\max}^{1/3}
    \label{eq:eta_c}
\end{align}

\begin{figure}[h!]
\begin{subfigure}{0.5\columnwidth}
        \centering
        \includegraphics[width=1.\linewidth]{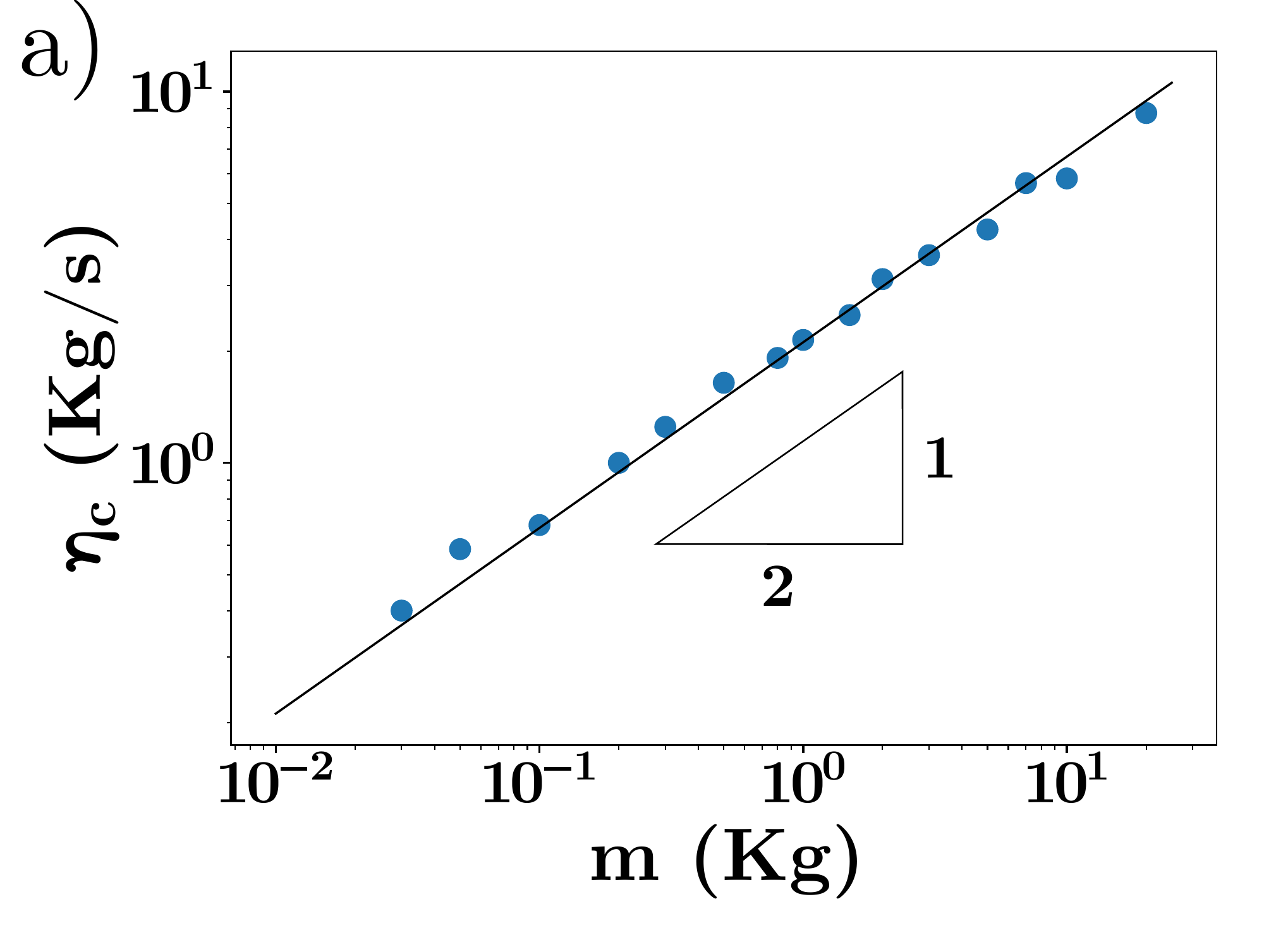}
\end{subfigure}%
    \hfill%
\begin{subfigure}{0.5\columnwidth}
        \centering
        \includegraphics[width=1.\linewidth]{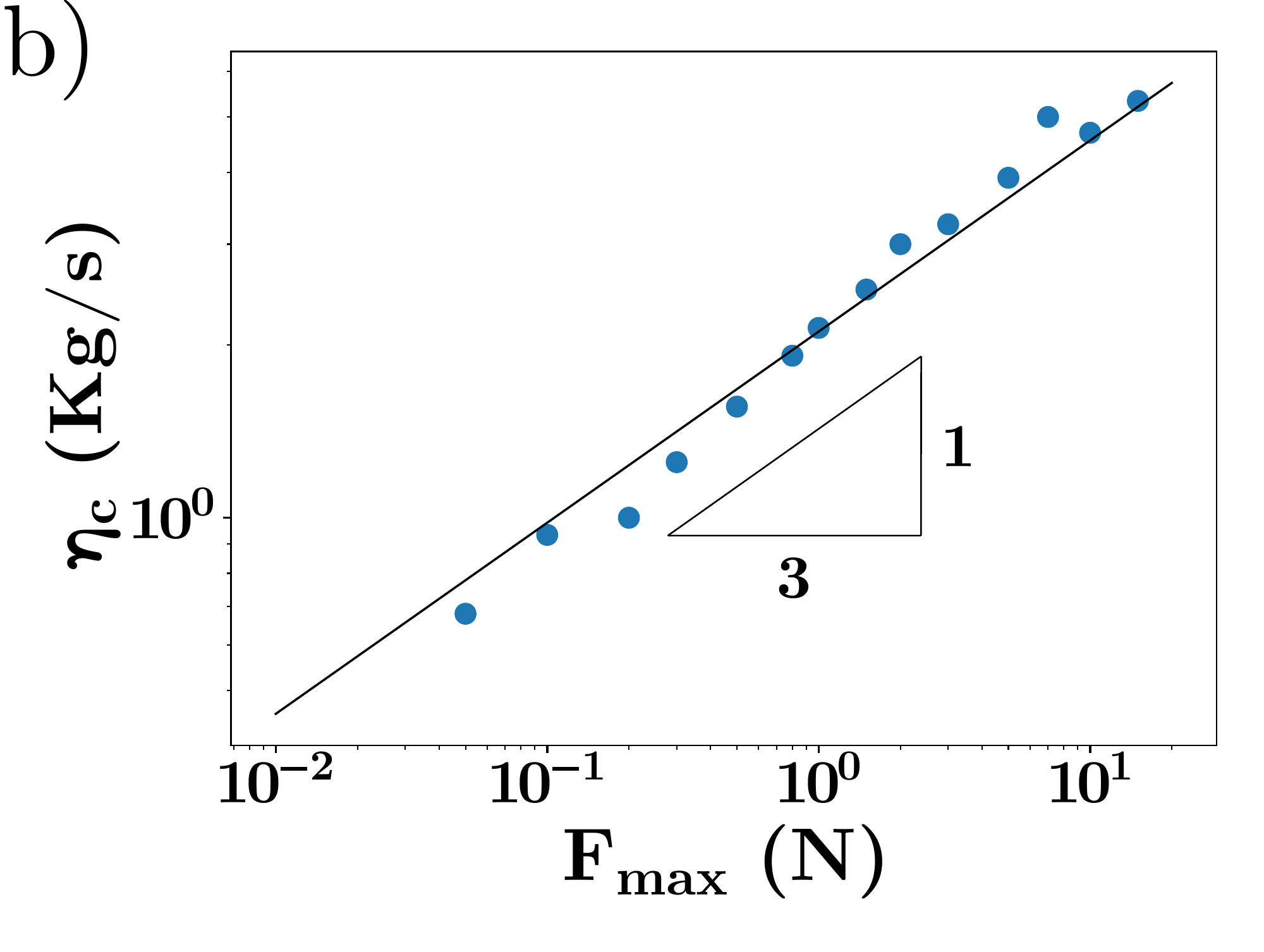}
\end{subfigure}%
    \caption{Evolution of the critical viscous coefficient $\eta_c$ with respect to a) the masses $m$ and b) the maximum amplitude of the external load $F_{\max}$. 
    The lines show the scaling behavior $\eta_c \propto m^{1/2} F_{\max}^{1/3}$.}
    \label{fig:scaling}
\end{figure}

To test this prediction, we investigate how the damping crossover $\eta_c$ observed in designed policies varies as the masses $m$ and the maximum force $F_{\max}$ are varied, exploring more than two orders of magnitude for both parameters. As show in Fig.\ref{fig:scaling}, the results present an excellent agreement with the proposed scaling argument~[\ref{eq:eta_c}].

\section*{Discussions and conclusion}
We have shown a proof of concept of general memory writing operations in a strongly non-linear system of coupled bi-stable springs by a reinforcement learning agent. 
In particular, we found that this technique allows reaching otherwise unreachable memory states dynamically. 
Interestingly, the agent appears to learn how to harness the physics underlying the behavior of the system: its control strategy changes qualitatively as the viscous coefficient is varied, from a relatively simple actuation in the large dissipation regime to a jerky dynamical behavior aimed at extracting the excess energy in the small dissipation regime. 
This transition coincides with a change in the system's internal response, from an over-damped to an inertial response.
As such, not only did it discover this transition by itself, pointing to physically relevant characteristic length scales, but it also signals an optimal design to achieve the most efficient memory manipulation.
In that sense, it was able to gather and share with the authors some insightful knowledge about the physics of the memory system, thus displaying some form of intelligence in {\it understanding} the challenges it was asked to tackle.  
Key stakes of future work will consist in identifying the cognitive structures established by the agent to complete the learned tasks, i.e. by rationalizing its neural activity and learning dynamics, using this knowledge to learn transitions for a higher number of coupled units.
Indeed, while we managed to successfully train networks on a system of four spring (see the {\it Materials and Methods} section), training on larger systems did not converge.
Finally, it would be interesting to verify whether the principles we discovered in this simulated model remain relevant in real-life situations.

\section*{Acknoledgements}
Research by T.J. was supported in part by the Raymond and Beverly Sackler Post-Doctoral Scholarship.

\bibliography{biblio}

\providecommand{\noopsort}[1]{}\providecommand{\singleletter}[1]{#1}%
\begin{thebibliography}{38}%
\makeatletter
\providecommand \@ifxundefined [1]{%
 \@ifx{#1\undefined}
}%
\providecommand \@ifnum [1]{%
 \ifnum #1\expandafter \@firstoftwo
 \else \expandafter \@secondoftwo
 \fi
}%
\providecommand \@ifx [1]{%
 \ifx #1\expandafter \@firstoftwo
 \else \expandafter \@secondoftwo
 \fi
}%
\providecommand \natexlab [1]{#1}%
\providecommand \enquote  [1]{``#1''}%
\providecommand \bibnamefont  [1]{#1}%
\providecommand \bibfnamefont [1]{#1}%
\providecommand \citenamefont [1]{#1}%
\providecommand \href@noop [0]{\@secondoftwo}%
\providecommand \href [0]{\begingroup \@sanitize@url \@href}%
\providecommand \@href[1]{\@@startlink{#1}\@@href}%
\providecommand \@@href[1]{\endgroup#1\@@endlink}%
\providecommand \@sanitize@url [0]{\catcode `\\12\catcode `\$12\catcode
  `\&12\catcode `\#12\catcode `\^12\catcode `\_12\catcode `\%12\relax}%
\providecommand \@@startlink[1]{}%
\providecommand \@@endlink[0]{}%
\providecommand \url  [0]{\begingroup\@sanitize@url \@url }%
\providecommand \@url [1]{\endgroup\@href {#1}{\urlprefix }}%
\providecommand \urlprefix  [0]{URL }%
\providecommand \Eprint [0]{\href }%
\providecommand \doibase [0]{https://doi.org/}%
\providecommand \selectlanguage [0]{\@gobble}%
\providecommand \bibinfo  [0]{\@secondoftwo}%
\providecommand \bibfield  [0]{\@secondoftwo}%
\providecommand \translation [1]{[#1]}%
\providecommand \BibitemOpen [0]{}%
\providecommand \bibitemStop [0]{}%
\providecommand \bibitemNoStop [0]{.\EOS\space}%
\providecommand \EOS [0]{\spacefactor3000\relax}%
\providecommand \BibitemShut  [1]{\csname bibitem#1\endcsname}%
\let\auto@bib@innerbib\@empty
\bibitem [{\citenamefont {Keim}\ \emph {et~al.}(2019)\citenamefont {Keim},
  \citenamefont {Paulsen}, \citenamefont {Zeravcic}, \citenamefont {Sastry},\
  and\ \citenamefont {Nagel}}]{keim2019memory}%
  \BibitemOpen
  \bibfield  {author} {\bibinfo {author} {\bibfnamefont {N.~C.}\ \bibnamefont
  {Keim}}, \bibinfo {author} {\bibfnamefont {J.~D.}\ \bibnamefont {Paulsen}},
  \bibinfo {author} {\bibfnamefont {Z.}~\bibnamefont {Zeravcic}}, \bibinfo
  {author} {\bibfnamefont {S.}~\bibnamefont {Sastry}},\ and\ \bibinfo {author}
  {\bibfnamefont {S.~R.}\ \bibnamefont {Nagel}},\ }\bibfield  {title} {\bibinfo
  {title} {Memory formation in matter},\ }\href@noop {} {\bibfield  {journal}
  {\bibinfo  {journal} {Reviews of Modern Physics}\ }\textbf {\bibinfo {volume}
  {91}},\ \bibinfo {pages} {035002} (\bibinfo {year} {2019})}\BibitemShut
  {NoStop}%
\bibitem [{\citenamefont {Kovacs}(1963)}]{kovacs1963glass}%
  \BibitemOpen
  \bibfield  {author} {\bibinfo {author} {\bibfnamefont {A.}~\bibnamefont
  {Kovacs}},\ }\bibfield  {title} {\bibinfo {title} {Glass transition in
  amorphous polymers: a phenomenological study},\ }\href@noop {} {\bibfield
  {journal} {\bibinfo  {journal} {Adv. Polym. Sci}\ }\textbf {\bibinfo {volume}
  {3}},\ \bibinfo {pages} {394} (\bibinfo {year} {1963})}\BibitemShut {NoStop}%
\bibitem [{\citenamefont {Prados}\ and\ \citenamefont
  {Trizac}(2014)}]{prados2014kovacs}%
  \BibitemOpen
  \bibfield  {author} {\bibinfo {author} {\bibfnamefont {A.}~\bibnamefont
  {Prados}}\ and\ \bibinfo {author} {\bibfnamefont {E.}~\bibnamefont
  {Trizac}},\ }\bibfield  {title} {\bibinfo {title} {Kovacs-like memory effect
  in driven granular gases},\ }\href@noop {} {\bibfield  {journal} {\bibinfo
  {journal} {Physical review letters}\ }\textbf {\bibinfo {volume} {112}},\
  \bibinfo {pages} {198001} (\bibinfo {year} {2014})}\BibitemShut {NoStop}%
\bibitem [{\citenamefont {Jules}\ \emph {et~al.}(2020)\citenamefont {Jules},
  \citenamefont {Lechenault},\ and\ \citenamefont
  {Adda-Bedia}}]{jules2020plasticity}%
  \BibitemOpen
  \bibfield  {author} {\bibinfo {author} {\bibfnamefont {T.}~\bibnamefont
  {Jules}}, \bibinfo {author} {\bibfnamefont {F.}~\bibnamefont {Lechenault}},\
  and\ \bibinfo {author} {\bibfnamefont {M.}~\bibnamefont {Adda-Bedia}},\
  }\bibfield  {title} {\bibinfo {title} {Plasticity and aging of folded elastic
  sheets},\ }\href@noop {} {\bibfield  {journal} {\bibinfo  {journal} {Physical
  Review E}\ }\textbf {\bibinfo {volume} {102}},\ \bibinfo {pages} {033005}
  (\bibinfo {year} {2020})}\BibitemShut {NoStop}%
\bibitem [{\citenamefont {Matan}\ \emph {et~al.}(2002)\citenamefont {Matan},
  \citenamefont {Williams}, \citenamefont {Witten},\ and\ \citenamefont
  {Nagel}}]{Matan2002}%
  \BibitemOpen
  \bibfield  {author} {\bibinfo {author} {\bibfnamefont {K.}~\bibnamefont
  {Matan}}, \bibinfo {author} {\bibfnamefont {R.~B.}\ \bibnamefont {Williams}},
  \bibinfo {author} {\bibfnamefont {T.~A.}\ \bibnamefont {Witten}},\ and\
  \bibinfo {author} {\bibfnamefont {S.~R.}\ \bibnamefont {Nagel}},\ }\bibfield
  {title} {\bibinfo {title} {Crumpling a thin sheet},\ }\href
  {https://doi.org/10.1103/PhysRevLett.88.076101} {\bibfield  {journal}
  {\bibinfo  {journal} {Phys. Rev. Lett.}\ }\textbf {\bibinfo {volume} {88}},\
  \bibinfo {pages} {076101} (\bibinfo {year} {2002})}\BibitemShut {NoStop}%
\bibitem [{\citenamefont {Diani}\ \emph {et~al.}(2009)\citenamefont {Diani},
  \citenamefont {Fayolle},\ and\ \citenamefont {Gilormini}}]{Diani2009}%
  \BibitemOpen
  \bibfield  {author} {\bibinfo {author} {\bibfnamefont {J.}~\bibnamefont
  {Diani}}, \bibinfo {author} {\bibfnamefont {B.}~\bibnamefont {Fayolle}},\
  and\ \bibinfo {author} {\bibfnamefont {P.}~\bibnamefont {Gilormini}},\
  }\bibfield  {title} {\bibinfo {title} {A review on the mullins effect},\
  }\href {https://doi.org/https://doi.org/10.1016/j.eurpolymj.2008.11.017}
  {\bibfield  {journal} {\bibinfo  {journal} {European Polymer Journal}\
  }\textbf {\bibinfo {volume} {45}},\ \bibinfo {pages} {601} (\bibinfo {year}
  {2009})}\BibitemShut {NoStop}%
\bibitem [{\citenamefont
  {Preisach}(1935{\natexlab{a}})}]{preisach1935magnetische}%
  \BibitemOpen
  \bibfield  {author} {\bibinfo {author} {\bibfnamefont {F.}~\bibnamefont
  {Preisach}},\ }\bibfield  {title} {\bibinfo {title} {{\"U}ber die magnetische
  nachwirkung},\ }\href@noop {} {\bibfield  {journal} {\bibinfo  {journal}
  {Zeitschrift f{\"u}r physik}\ }\textbf {\bibinfo {volume} {94}},\ \bibinfo
  {pages} {277} (\bibinfo {year} {1935}{\natexlab{a}})}\BibitemShut {NoStop}%
\bibitem [{\citenamefont {Mayergoyz}(1986)}]{mayergoyz1986mathematical}%
  \BibitemOpen
  \bibfield  {author} {\bibinfo {author} {\bibfnamefont {I.}~\bibnamefont
  {Mayergoyz}},\ }\bibfield  {title} {\bibinfo {title} {Mathematical models of
  hysteresis},\ }\href@noop {} {\bibfield  {journal} {\bibinfo  {journal}
  {Physical Review Letters}\ }\textbf {\bibinfo {volume} {56}},\ \bibinfo
  {pages} {1518} (\bibinfo {year} {1986})}\BibitemShut {NoStop}%
\bibitem [{\citenamefont {John Adair~Barker}(1983)}]{Barker1983a}%
  \BibitemOpen
  \bibfield  {author} {\bibinfo {author} {\bibfnamefont {B.~G. H. D. H.~E.}\
  \bibnamefont {John Adair~Barker}, \bibfnamefont {D.~E.~Schreiber}},\
  }\bibfield  {title} {\bibinfo {title} {Magnetic hysteresis and minor loops:
  models and experiments},\ }\href {https://doi.org/10.1098/rspa.1983.0035}
  {\bibfield  {journal} {\bibinfo  {journal} {Proceedings of the Royal Society
  of London. A. Mathematical and Physical Sciences}\ }\textbf {\bibinfo
  {volume} {386}},\ \bibinfo {pages} {251} (\bibinfo {year}
  {1983})}\BibitemShut {NoStop}%
\bibitem [{\citenamefont {Deutsch}\ \emph {et~al.}(2004)\citenamefont
  {Deutsch}, \citenamefont {Dhar},\ and\ \citenamefont
  {Narayan}}]{Deutsch2004}%
  \BibitemOpen
  \bibfield  {author} {\bibinfo {author} {\bibfnamefont {J.~M.}\ \bibnamefont
  {Deutsch}}, \bibinfo {author} {\bibfnamefont {A.}~\bibnamefont {Dhar}},\ and\
  \bibinfo {author} {\bibfnamefont {O.}~\bibnamefont {Narayan}},\ }\bibfield
  {title} {\bibinfo {title} {Return to return point memory},\ }\href
  {https://doi.org/10.1103/PhysRevLett.92.227203} {\bibfield  {journal}
  {\bibinfo  {journal} {Phys. Rev. Lett.}\ }\textbf {\bibinfo {volume} {92}},\
  \bibinfo {pages} {227203} (\bibinfo {year} {2004})}\BibitemShut {NoStop}%
\bibitem [{\citenamefont {Mungan}\ and\ \citenamefont
  {Terzi}(2019)}]{mungan2019structure}%
  \BibitemOpen
  \bibfield  {author} {\bibinfo {author} {\bibfnamefont {M.}~\bibnamefont
  {Mungan}}\ and\ \bibinfo {author} {\bibfnamefont {M.~M.}\ \bibnamefont
  {Terzi}},\ }\bibfield  {title} {\bibinfo {title} {The structure of state
  transition graphs in systems with return point memory: I. general theory},\
  }in\ \href@noop {} {\emph {\bibinfo {booktitle} {Annales Henri
  Poincar{\'e}}}},\ Vol.~\bibinfo {volume} {20}\ (\bibinfo {organization}
  {Springer},\ \bibinfo {year} {2019})\ pp.\ \bibinfo {pages}
  {2819--2872}\BibitemShut {NoStop}%
\bibitem [{\citenamefont {Keim}\ \emph {et~al.}(2020)\citenamefont {Keim},
  \citenamefont {Hass}, \citenamefont {Kroger},\ and\ \citenamefont
  {Wieker}}]{keim2020global}%
  \BibitemOpen
  \bibfield  {author} {\bibinfo {author} {\bibfnamefont {N.~C.}\ \bibnamefont
  {Keim}}, \bibinfo {author} {\bibfnamefont {J.}~\bibnamefont {Hass}}, \bibinfo
  {author} {\bibfnamefont {B.}~\bibnamefont {Kroger}},\ and\ \bibinfo {author}
  {\bibfnamefont {D.}~\bibnamefont {Wieker}},\ }\bibfield  {title} {\bibinfo
  {title} {Global memory from local hysteresis in an amorphous solid},\
  }\href@noop {} {\bibfield  {journal} {\bibinfo  {journal} {Physical Review
  Research}\ }\textbf {\bibinfo {volume} {2}},\ \bibinfo {pages} {012004}
  (\bibinfo {year} {2020})}\BibitemShut {NoStop}%
\bibitem [{\citenamefont {Keim}\ and\ \citenamefont
  {Paulsen}(2021)}]{Keim2021a}%
  \BibitemOpen
  \bibfield  {author} {\bibinfo {author} {\bibfnamefont {N.~C.}\ \bibnamefont
  {Keim}}\ and\ \bibinfo {author} {\bibfnamefont {J.~D.}\ \bibnamefont
  {Paulsen}},\ }\bibfield  {title} {\bibinfo {title} {Multiperiodic orbits from
  interacting soft spots in cyclically sheared amorphous solids},\ }\href
  {https://doi.org/10.1126/sciadv.abg7685} {\bibfield  {journal} {\bibinfo
  {journal} {Science Advances}\ }\textbf {\bibinfo {volume} {7}},\ \bibinfo
  {pages} {eabg7685} (\bibinfo {year} {2021})},\ \Eprint
  {https://arxiv.org/abs/https://www.science.org/doi/pdf/10.1126/sciadv.abg7685}
  {https://www.science.org/doi/pdf/10.1126/sciadv.abg7685} \BibitemShut
  {NoStop}%
\bibitem [{\citenamefont {Lib{\'a}l}\ \emph {et~al.}(2012)\citenamefont
  {Lib{\'a}l}, \citenamefont {Reichhardt},\ and\ \citenamefont
  {Reichhardt}}]{libal2012hysteresis}%
  \BibitemOpen
  \bibfield  {author} {\bibinfo {author} {\bibfnamefont {A.}~\bibnamefont
  {Lib{\'a}l}}, \bibinfo {author} {\bibfnamefont {C.}~\bibnamefont
  {Reichhardt}},\ and\ \bibinfo {author} {\bibfnamefont {C.~O.}\ \bibnamefont
  {Reichhardt}},\ }\bibfield  {title} {\bibinfo {title} {Hysteresis and
  return-point memory in colloidal artificial spin ice systems},\ }\href@noop
  {} {\bibfield  {journal} {\bibinfo  {journal} {Physical Review E}\ }\textbf
  {\bibinfo {volume} {86}},\ \bibinfo {pages} {021406} (\bibinfo {year}
  {2012})}\BibitemShut {NoStop}%
\bibitem [{\citenamefont {Goicoechea}\ and\ \citenamefont
  {Ort{\'\i}n}(1994)}]{goicoechea1994hysteresis}%
  \BibitemOpen
  \bibfield  {author} {\bibinfo {author} {\bibfnamefont {J.}~\bibnamefont
  {Goicoechea}}\ and\ \bibinfo {author} {\bibfnamefont {J.}~\bibnamefont
  {Ort{\'\i}n}},\ }\bibfield  {title} {\bibinfo {title} {Hysteresis and
  return-point memory in deterministic cellular automata},\ }\href@noop {}
  {\bibfield  {journal} {\bibinfo  {journal} {Physical review letters}\
  }\textbf {\bibinfo {volume} {72}},\ \bibinfo {pages} {2203} (\bibinfo {year}
  {1994})}\BibitemShut {NoStop}%
\bibitem [{\citenamefont {Bense}\ and\ \citenamefont {van
  Hecke}(2021)}]{bense2021complex}%
  \BibitemOpen
  \bibfield  {author} {\bibinfo {author} {\bibfnamefont {H.}~\bibnamefont
  {Bense}}\ and\ \bibinfo {author} {\bibfnamefont {M.}~\bibnamefont {van
  Hecke}},\ }\bibfield  {title} {\bibinfo {title} {Complex pathways and memory
  in compressed corrugated sheets},\ }\href@noop {} {\bibfield  {journal}
  {\bibinfo  {journal} {Proceedings of the National Academy of Sciences}\
  }\textbf {\bibinfo {volume} {118}} (\bibinfo {year} {2021})}\BibitemShut
  {NoStop}%
\bibitem [{\citenamefont {Lindeman}\ and\ \citenamefont
  {Nagel}(2021)}]{lindeman2021multiple}%
  \BibitemOpen
  \bibfield  {author} {\bibinfo {author} {\bibfnamefont {C.~W.}\ \bibnamefont
  {Lindeman}}\ and\ \bibinfo {author} {\bibfnamefont {S.~R.}\ \bibnamefont
  {Nagel}},\ }\bibfield  {title} {\bibinfo {title} {Multiple memory formation
  in glassy landscapes},\ }\href@noop {} {\bibfield  {journal} {\bibinfo
  {journal} {Science Advances}\ }\textbf {\bibinfo {volume} {7}},\ \bibinfo
  {pages} {eabg7133} (\bibinfo {year} {2021})}\BibitemShut {NoStop}%
\bibitem [{\citenamefont {Puglisi}\ and\ \citenamefont
  {Truskinovsky}(2002{\natexlab{a}})}]{puglisi2002mechanism}%
  \BibitemOpen
  \bibfield  {author} {\bibinfo {author} {\bibfnamefont {G.}~\bibnamefont
  {Puglisi}}\ and\ \bibinfo {author} {\bibfnamefont {L.}~\bibnamefont
  {Truskinovsky}},\ }\bibfield  {title} {\bibinfo {title} {A mechanism of
  transformational plasticity},\ }\href@noop {} {\bibfield  {journal} {\bibinfo
   {journal} {Continuum Mechanics and Thermodynamics}\ }\textbf {\bibinfo
  {volume} {14}},\ \bibinfo {pages} {437} (\bibinfo {year}
  {2002}{\natexlab{a}})}\BibitemShut {NoStop}%
\bibitem [{\citenamefont {Regev}\ \emph {et~al.}(2021)\citenamefont {Regev},
  \citenamefont {Attia}, \citenamefont {Dahmen}, \citenamefont {Sastry},\ and\
  \citenamefont {Mungan}}]{regev2021topology}%
  \BibitemOpen
  \bibfield  {author} {\bibinfo {author} {\bibfnamefont {I.}~\bibnamefont
  {Regev}}, \bibinfo {author} {\bibfnamefont {I.}~\bibnamefont {Attia}},
  \bibinfo {author} {\bibfnamefont {K.}~\bibnamefont {Dahmen}}, \bibinfo
  {author} {\bibfnamefont {S.}~\bibnamefont {Sastry}},\ and\ \bibinfo {author}
  {\bibfnamefont {M.}~\bibnamefont {Mungan}},\ }\bibfield  {title} {\bibinfo
  {title} {Topology of the energy landscape of sheared amorphous solids and the
  irreversibility transition},\ }\href@noop {} {\bibfield  {journal} {\bibinfo
  {journal} {Physical Review E}\ }\textbf {\bibinfo {volume} {103}},\ \bibinfo
  {pages} {062614} (\bibinfo {year} {2021})}\BibitemShut {NoStop}%
\bibitem [{\citenamefont {Yasuda}\ \emph {et~al.}(2017)\citenamefont {Yasuda},
  \citenamefont {Tachi}, \citenamefont {Lee},\ and\ \citenamefont
  {Yang}}]{yasuda2017origami}%
  \BibitemOpen
  \bibfield  {author} {\bibinfo {author} {\bibfnamefont {H.}~\bibnamefont
  {Yasuda}}, \bibinfo {author} {\bibfnamefont {T.}~\bibnamefont {Tachi}},
  \bibinfo {author} {\bibfnamefont {M.}~\bibnamefont {Lee}},\ and\ \bibinfo
  {author} {\bibfnamefont {J.}~\bibnamefont {Yang}},\ }\bibfield  {title}
  {\bibinfo {title} {Origami-based tunable truss structures for non-volatile
  mechanical memory operation},\ }\href@noop {} {\bibfield  {journal} {\bibinfo
   {journal} {Nature communications}\ }\textbf {\bibinfo {volume} {8}},\
  \bibinfo {pages} {1} (\bibinfo {year} {2017})}\BibitemShut {NoStop}%
\bibitem [{\citenamefont {Jules}\ \emph {et~al.}(2022)\citenamefont {Jules},
  \citenamefont {Reid}, \citenamefont {Daniels}, \citenamefont {Mungan},\ and\
  \citenamefont {Lechenault}}]{jules2022delicate}%
  \BibitemOpen
  \bibfield  {author} {\bibinfo {author} {\bibfnamefont {T.}~\bibnamefont
  {Jules}}, \bibinfo {author} {\bibfnamefont {A.}~\bibnamefont {Reid}},
  \bibinfo {author} {\bibfnamefont {K.~E.}\ \bibnamefont {Daniels}}, \bibinfo
  {author} {\bibfnamefont {M.}~\bibnamefont {Mungan}},\ and\ \bibinfo {author}
  {\bibfnamefont {F.}~\bibnamefont {Lechenault}},\ }\bibfield  {title}
  {\bibinfo {title} {Delicate memory structure of origami switches},\
  }\href@noop {} {\bibfield  {journal} {\bibinfo  {journal} {Physical Review
  Research}\ }\textbf {\bibinfo {volume} {4}},\ \bibinfo {pages} {013128}
  (\bibinfo {year} {2022})}\BibitemShut {NoStop}%
\bibitem [{\citenamefont {Terzi}\ and\ \citenamefont
  {Mungan}(2020)}]{terzi2020state}%
  \BibitemOpen
  \bibfield  {author} {\bibinfo {author} {\bibfnamefont {M.~M.}\ \bibnamefont
  {Terzi}}\ and\ \bibinfo {author} {\bibfnamefont {M.}~\bibnamefont {Mungan}},\
  }\bibfield  {title} {\bibinfo {title} {State transition graph of the preisach
  model and the role of return-point memory},\ }\href
  {https://doi.org/10.1103/PhysRevE.102.012122} {\bibfield  {journal} {\bibinfo
   {journal} {Phys. Rev. E}\ }\textbf {\bibinfo {volume} {102}},\ \bibinfo
  {pages} {012122} (\bibinfo {year} {2020})}\BibitemShut {NoStop}%
\bibitem [{\citenamefont {van Hecke}(2021)}]{VanHecke2021a}%
  \BibitemOpen
  \bibfield  {author} {\bibinfo {author} {\bibfnamefont {M.}~\bibnamefont {van
  Hecke}},\ }\bibfield  {title} {\bibinfo {title} {Profusion of transition
  pathways for interacting hysterons},\ }\href
  {https://doi.org/10.1103/PhysRevE.104.054608} {\bibfield  {journal} {\bibinfo
   {journal} {Phys. Rev. E}\ }\textbf {\bibinfo {volume} {104}},\ \bibinfo
  {pages} {054608} (\bibinfo {year} {2021})}\BibitemShut {NoStop}%
\bibitem [{\citenamefont {Puglisi}\ and\ \citenamefont
  {Truskinovsky}(2002{\natexlab{b}})}]{puglisi2002rate}%
  \BibitemOpen
  \bibfield  {author} {\bibinfo {author} {\bibfnamefont {G.}~\bibnamefont
  {Puglisi}}\ and\ \bibinfo {author} {\bibfnamefont {L.}~\bibnamefont
  {Truskinovsky}},\ }\bibfield  {title} {\bibinfo {title} {Rate independent
  hysteresis in a bi-stable chain},\ }\href@noop {} {\bibfield  {journal}
  {\bibinfo  {journal} {Journal of the Mechanics and Physics of Solids}\
  }\textbf {\bibinfo {volume} {50}},\ \bibinfo {pages} {165} (\bibinfo {year}
  {2002}{\natexlab{b}})}\BibitemShut {NoStop}%
\bibitem [{\citenamefont {Pan}\ and\ \citenamefont
  {Yang}(2009)}]{pan2009survey}%
  \BibitemOpen
  \bibfield  {author} {\bibinfo {author} {\bibfnamefont {S.~J.}\ \bibnamefont
  {Pan}}\ and\ \bibinfo {author} {\bibfnamefont {Q.}~\bibnamefont {Yang}},\
  }\bibfield  {title} {\bibinfo {title} {A survey on transfer learning},\
  }\href@noop {} {\bibfield  {journal} {\bibinfo  {journal} {IEEE Transactions
  on knowledge and data engineering}\ }\textbf {\bibinfo {volume} {22}},\
  \bibinfo {pages} {1345} (\bibinfo {year} {2009})}\BibitemShut {NoStop}%
\bibitem [{\citenamefont {Taylor}\ and\ \citenamefont
  {Stone}(2009)}]{taylor2009transfer}%
  \BibitemOpen
  \bibfield  {author} {\bibinfo {author} {\bibfnamefont {M.~E.}\ \bibnamefont
  {Taylor}}\ and\ \bibinfo {author} {\bibfnamefont {P.}~\bibnamefont {Stone}},\
  }\bibfield  {title} {\bibinfo {title} {Transfer learning for reinforcement
  learning domains: A survey.},\ }\href@noop {} {\bibfield  {journal} {\bibinfo
   {journal} {Journal of Machine Learning Research}\ }\textbf {\bibinfo
  {volume} {10}} (\bibinfo {year} {2009})}\BibitemShut {NoStop}%
\bibitem [{\citenamefont {Preisach}(1935{\natexlab{b}})}]{Preisach1935}%
  \BibitemOpen
  \bibfield  {author} {\bibinfo {author} {\bibfnamefont {F.}~\bibnamefont
  {Preisach}},\ }\bibfield  {title} {\bibinfo {title} {{\"Uber die magnetische
  Nachwirkung}},\ }\href {https://doi.org/10.1007/bf01349418} {\bibfield
  {journal} {\bibinfo  {journal} {Zeitschrift f\"ur Physik}\ }\textbf {\bibinfo
  {volume} {94}},\ \bibinfo {pages} {277} (\bibinfo {year}
  {1935}{\natexlab{b}})}\BibitemShut {NoStop}%
\bibitem [{\citenamefont {Gadaleta}\ and\ \citenamefont
  {Dangelmayr}(1999)}]{gadaleta1999optimal}%
  \BibitemOpen
  \bibfield  {author} {\bibinfo {author} {\bibfnamefont {S.}~\bibnamefont
  {Gadaleta}}\ and\ \bibinfo {author} {\bibfnamefont {G.}~\bibnamefont
  {Dangelmayr}},\ }\bibfield  {title} {\bibinfo {title} {Optimal chaos control
  through reinforcement learning},\ }\href@noop {} {\bibfield  {journal}
  {\bibinfo  {journal} {Chaos: An Interdisciplinary Journal of Nonlinear
  Science}\ }\textbf {\bibinfo {volume} {9}},\ \bibinfo {pages} {775} (\bibinfo
  {year} {1999})}\BibitemShut {NoStop}%
\bibitem [{\citenamefont {Gadaleta}\ and\ \citenamefont
  {Dangelmayr}(2001)}]{gadaleta2001learning}%
  \BibitemOpen
  \bibfield  {author} {\bibinfo {author} {\bibfnamefont {S.}~\bibnamefont
  {Gadaleta}}\ and\ \bibinfo {author} {\bibfnamefont {G.}~\bibnamefont
  {Dangelmayr}},\ }\bibfield  {title} {\bibinfo {title} {Learning to control a
  complex multistable system},\ }\href@noop {} {\bibfield  {journal} {\bibinfo
  {journal} {Physical Review E}\ }\textbf {\bibinfo {volume} {63}},\ \bibinfo
  {pages} {036217} (\bibinfo {year} {2001})}\BibitemShut {NoStop}%
\bibitem [{\citenamefont {Wang}\ \emph {et~al.}(2021)\citenamefont {Wang},
  \citenamefont {Turner},\ and\ \citenamefont {Mann}}]{wang2021constrained}%
  \BibitemOpen
  \bibfield  {author} {\bibinfo {author} {\bibfnamefont {X.-S.}\ \bibnamefont
  {Wang}}, \bibinfo {author} {\bibfnamefont {J.~D.}\ \bibnamefont {Turner}},\
  and\ \bibinfo {author} {\bibfnamefont {B.~P.}\ \bibnamefont {Mann}},\
  }\bibfield  {title} {\bibinfo {title} {Constrained attractor selection using
  deep reinforcement learning},\ }\href@noop {} {\bibfield  {journal} {\bibinfo
   {journal} {Journal of Vibration and Control}\ }\textbf {\bibinfo {volume}
  {27}},\ \bibinfo {pages} {502} (\bibinfo {year} {2021})}\BibitemShut
  {NoStop}%
\bibitem [{\citenamefont {Pisarchik}\ and\ \citenamefont
  {Feudel}(2014)}]{pisarchik2014control}%
  \BibitemOpen
  \bibfield  {author} {\bibinfo {author} {\bibfnamefont {A.~N.}\ \bibnamefont
  {Pisarchik}}\ and\ \bibinfo {author} {\bibfnamefont {U.}~\bibnamefont
  {Feudel}},\ }\bibfield  {title} {\bibinfo {title} {Control of
  multistability},\ }\href@noop {} {\bibfield  {journal} {\bibinfo  {journal}
  {Physics Reports}\ }\textbf {\bibinfo {volume} {540}},\ \bibinfo {pages}
  {167} (\bibinfo {year} {2014})}\BibitemShut {NoStop}%
\bibitem [{\citenamefont {Konda}\ and\ \citenamefont
  {Tsitsiklis}(1999)}]{konda1999actor}%
  \BibitemOpen
  \bibfield  {author} {\bibinfo {author} {\bibfnamefont {V.}~\bibnamefont
  {Konda}}\ and\ \bibinfo {author} {\bibfnamefont {J.}~\bibnamefont
  {Tsitsiklis}},\ }\bibfield  {title} {\bibinfo {title} {Actor-critic
  algorithms},\ }\href@noop {} {\bibfield  {journal} {\bibinfo  {journal}
  {Advances in neural information processing systems}\ }\textbf {\bibinfo
  {volume} {12}} (\bibinfo {year} {1999})}\BibitemShut {NoStop}%
\bibitem [{\citenamefont {Grondman}\ \emph {et~al.}(2011)\citenamefont
  {Grondman}, \citenamefont {Vaandrager}, \citenamefont {Busoniu},
  \citenamefont {Babuska},\ and\ \citenamefont
  {Schuitema}}]{grondman2011efficient}%
  \BibitemOpen
  \bibfield  {author} {\bibinfo {author} {\bibfnamefont {I.}~\bibnamefont
  {Grondman}}, \bibinfo {author} {\bibfnamefont {M.}~\bibnamefont
  {Vaandrager}}, \bibinfo {author} {\bibfnamefont {L.}~\bibnamefont {Busoniu}},
  \bibinfo {author} {\bibfnamefont {R.}~\bibnamefont {Babuska}},\ and\ \bibinfo
  {author} {\bibfnamefont {E.}~\bibnamefont {Schuitema}},\ }\bibfield  {title}
  {\bibinfo {title} {Efficient model learning methods for actor--critic
  control},\ }\href@noop {} {\bibfield  {journal} {\bibinfo  {journal} {IEEE
  Transactions on Systems, Man, and Cybernetics, Part B (Cybernetics)}\
  }\textbf {\bibinfo {volume} {42}},\ \bibinfo {pages} {591} (\bibinfo {year}
  {2011})}\BibitemShut {NoStop}%
\bibitem [{\citenamefont {Fujimoto}\ \emph {et~al.}(2018)\citenamefont
  {Fujimoto}, \citenamefont {Hoof},\ and\ \citenamefont {Meger}}]{TD3}%
  \BibitemOpen
  \bibfield  {author} {\bibinfo {author} {\bibfnamefont {S.}~\bibnamefont
  {Fujimoto}}, \bibinfo {author} {\bibfnamefont {H.}~\bibnamefont {Hoof}},\
  and\ \bibinfo {author} {\bibfnamefont {D.}~\bibnamefont {Meger}},\ }\bibfield
   {title} {\bibinfo {title} {Addressing function approximation error in
  actor-critic methods},\ }in\ \href@noop {} {\emph {\bibinfo {booktitle}
  {International conference on machine learning}}}\ (\bibinfo {organization}
  {PMLR},\ \bibinfo {year} {2018})\ pp.\ \bibinfo {pages}
  {1587--1596}\BibitemShut {NoStop}%
\bibitem [{\citenamefont {Brockman}\ \emph {et~al.}(2016)\citenamefont
  {Brockman}, \citenamefont {Cheung}, \citenamefont {Pettersson}, \citenamefont
  {Schneider}, \citenamefont {Tang},\ and\ \citenamefont {Zaremba}}]{gym}%
  \BibitemOpen
  \bibfield  {author} {\bibinfo {author} {\bibfnamefont {G.}~\bibnamefont
  {Brockman}}, \bibinfo {author} {\bibfnamefont {V.}~\bibnamefont {Cheung}},
  \bibinfo {author} {\bibfnamefont {L.}~\bibnamefont {Pettersson}}, \bibinfo
  {author} {\bibfnamefont {J.}~\bibnamefont {Schneider}}, \bibinfo {author}
  {\bibfnamefont {J.~S.~J.}\ \bibnamefont {Tang}},\ and\ \bibinfo {author}
  {\bibfnamefont {W.}~\bibnamefont {Zaremba}},\ }\bibfield  {title} {\bibinfo
  {title} {Openai gym},\ }\href@noop {} {\bibfield  {journal} {\bibinfo
  {journal} {arXiv preprint arXiv:1606.01540.}\ } (\bibinfo {year}
  {2016})}\BibitemShut {NoStop}%
\bibitem [{\citenamefont {Fujita}\ \emph {et~al.}(2021)\citenamefont {Fujita},
  \citenamefont {Nagarajan}, \citenamefont {Kataoka},\ and\ \citenamefont
  {Ishikawa}}]{JMLR:v22:20-376}%
  \BibitemOpen
  \bibfield  {author} {\bibinfo {author} {\bibfnamefont {Y.}~\bibnamefont
  {Fujita}}, \bibinfo {author} {\bibfnamefont {P.}~\bibnamefont {Nagarajan}},
  \bibinfo {author} {\bibfnamefont {T.}~\bibnamefont {Kataoka}},\ and\ \bibinfo
  {author} {\bibfnamefont {T.}~\bibnamefont {Ishikawa}},\ }\bibfield  {title}
  {\bibinfo {title} {Chainerrl: A deep reinforcement learning library},\ }\href
  {http://jmlr.org/papers/v22/20-376.html} {\bibfield  {journal} {\bibinfo
  {journal} {Journal of Machine Learning Research}\ }\textbf {\bibinfo {volume}
  {22}},\ \bibinfo {pages} {1} (\bibinfo {year} {2021})}\BibitemShut {NoStop}%
\bibitem [{\citenamefont {Kingma}\ and\ \citenamefont
  {Ba}(2014)}]{kingma2014adam}%
  \BibitemOpen
  \bibfield  {author} {\bibinfo {author} {\bibfnamefont {D.~P.}\ \bibnamefont
  {Kingma}}\ and\ \bibinfo {author} {\bibfnamefont {J.}~\bibnamefont {Ba}},\
  }\bibfield  {title} {\bibinfo {title} {Adam: A method for stochastic
  optimization},\ }\href@noop {} {\bibfield  {journal} {\bibinfo  {journal}
  {arXiv preprint arXiv:1412.6980}\ } (\bibinfo {year} {2014})}\BibitemShut
  {NoStop}%
\bibitem [{\citenamefont {Michel}\ \emph {et~al.}(2022)\citenamefont {Michel},
  \citenamefont {Jules},\ and\ \citenamefont {Douin}}]{RepoGit}%
  \BibitemOpen
  \bibfield  {author} {\bibinfo {author} {\bibfnamefont {L.}~\bibnamefont
  {Michel}}, \bibinfo {author} {\bibfnamefont {T.}~\bibnamefont {Jules}},\ and\
  \bibinfo {author} {\bibfnamefont {A.}~\bibnamefont {Douin}},\ }\href
  {https://doi.org/10.5281/zenodo.6514157} {\bibinfo {title}
  {{laura042/Multistable$\_$memory$\_$system: v0}}},\ \bibinfo {howpublished}
  {\url{https://doi.org/10.5281/zenodo.6514157}} (\bibinfo {year}
  {2022})\BibitemShut {NoStop}%
\end{thebibliography}%

\appendix

\section*{Description of our Reinforcement Learning setup}

\subsection*{Deep Deterministic Policy Gradient}

\begin{figure}[h!]
    \centering
    \includegraphics[width=1.\linewidth]{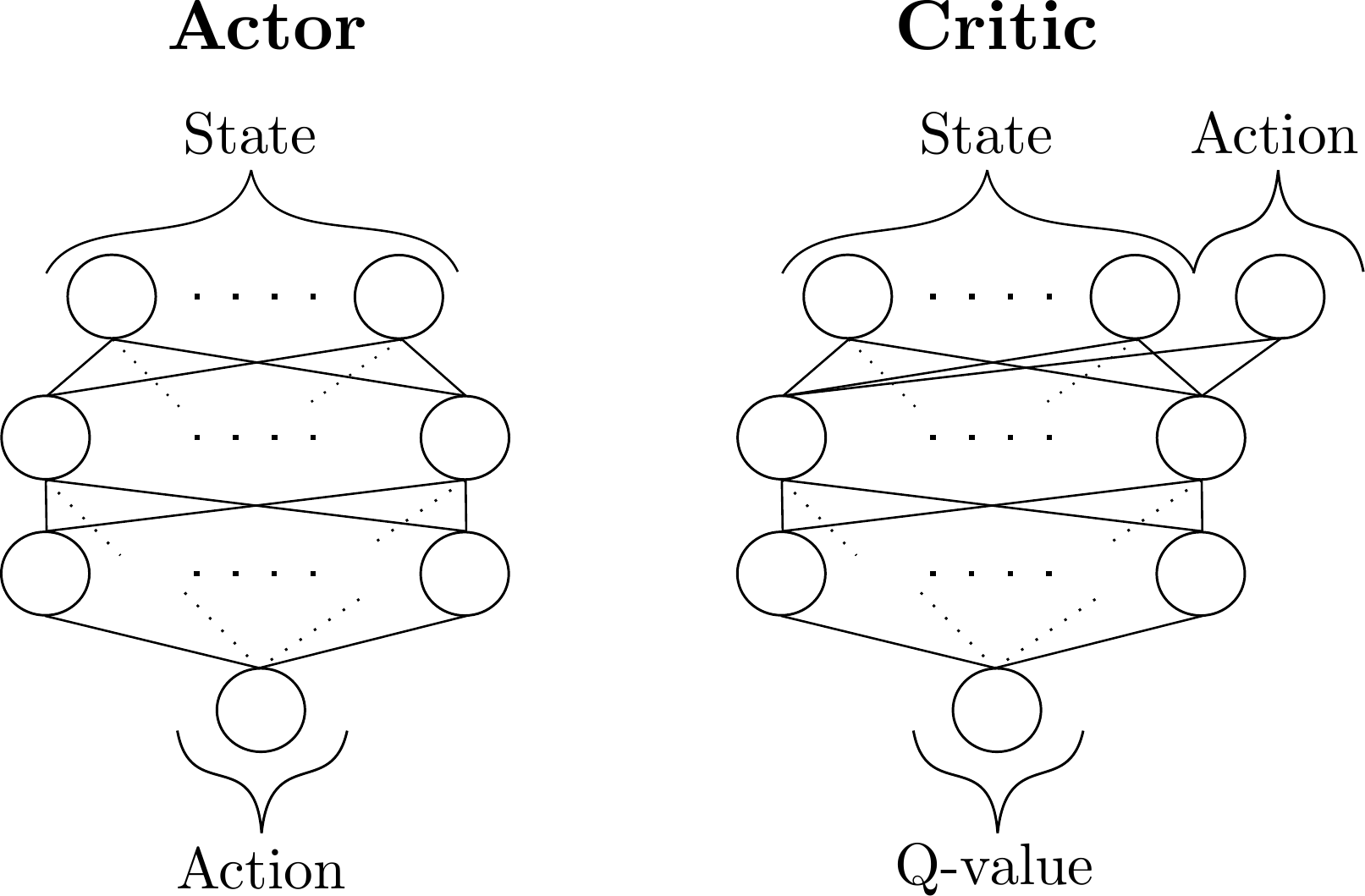}
    \caption{Actor and critic architectures.}
    \label{fig:actor_critic}
\end{figure}

The aim of a RL agent is to choose an action that will allow it to reach a specific goal in the future. Many RL algorithms learn an approximated Q-function in order to have access to the optimal actions to choose to reach the goal. The Q-function is the expectation of the future (discounted) reward for every action available to the agent given the state of the environment. The output of the Q-function is called the Q-value.
\begin{align}
    Q(s, a) = \mathbb{E}[R_{t}|a=a_{t}, s=s_{t}]
    \label{eq:qfunc}
\end{align}

with $R_{t} = \sum_{t'=t}^{T}\gamma^{t'-t}r_{t'}$ and $\gamma$ being a discount factor between 0 and 1. \\
The optimal Q-function satisfies Bellman equation.
\begin{equation}
    Q^*(s, a) = \mathbb{E}[r + \gamma\max_{a'}Q^*(s', a')|s, a]
\end{equation}
where $r$ is the reward obtained if the agent takes the action $a$ while the state of the environment is $s$, $a'$ and $s'$ are respectively the next action and the next state. \\
There is a simple relation between the optimal Q-function and the optimal action/policy.
\begin{align}
    a^*(s) = arg\max_{a}Q^*(s, a)
    \label{eq:optaction}
\end{align}
In Q-Learning algorithms, an iterative, self-consistent scheme is written to approximate Bellman's equation. The algorithm learns this approximation on the accumulated memory of its previous trials.
\begin{align}
    Q_{i+1}(s, a) = E[r + \gamma\max_{a'}Q_{i}(s', a')|s, a]
    \label{eq:bellman}
\end{align}
However, in our case, the agent chooses its actions in a continuous interval, creating an infinite number of possible actions and making equation [\ref{eq:optaction}] impractical. The Deep Deterministic Policy Gradient (DDPG) agent we chose addresses this problem by separately learning a Q-function and a policy. It approximates the optimal Q-function by a neural network, called the critic, and approximates the optimal policy (i. e. the optimal action) by a second neural network, called the actor  (see Fig.~\ref{fig:actor_critic}). The critic takes as an input the environment's state and the action chosen by the actor and outputs a Q-value. The actor takes the environment's state as an input and outputs a selected action. While playing, the agent stacks its experiences into a replay buffer and uses them randomly to update the Q-function and the policy. \\

\subsection*{Twin-Delayed DDPG}
The DDPG agent learns the optimal Q-function and the optimal policy concurrently. The policy indeed appears in the loss used to learn the model approximating the Q-function, and the Q-function appears in the loss of the neural network approximating the policy. An unwanted drawback is that the errors in the learned Q-function propagate to the policy. We then preferred to use the so-called Twin-Delayed DDPG~\cite{TD3} that addresses this issue. Twin-Delayed DDPG learns two Q-functions and uses the smaller of the two in the updates of its Q-functions. Noise is also added to the policy used in the Q-function loss, and the latter is updated less frequently than the Q-function. These additional mechanisms improve performance by limiting over-estimations and reducing per-update errors.

\subsection*{Gym package}

Gym is a tool developed by OpenAI\textcircled{c} to deploy reinforcement learning algorithms~\cite{gym}. Gym offers a wide variety of challenging environments and allows to easily create custom ones. We then used the Gym package to create the environment for our agent. In Gym formalism, training is divided into episodes themselves divided into steps. A step corresponds to each time the agent chooses an action. At the beginning of each episode, the state of the environment is randomly initialized and the agent possesses a defined number of steps to reach the goal of the game.

\section*{Description of our setup}

We used the Twin-Delayed DDPG agent implemented in~\cite{JMLR:v22:20-376} to solve our control problem. The task involves reaching a stable configuration close to rest, starting from random initial conditions. At the beginning of an episode, a target state is randomly chosen and the initial positions and velocities are randomly sampled from the respective intervals [$\delta_{i}^{(0)}$ - 0.2, $\delta_{i}^{(1)}$ + 0.2] and [-0.1, 0.1]. We begin training with random initial policy and Q-functions parameters. The weights and biases of each of layer are sampled from $U(-\sqrt{k}, \sqrt{k})$ where $U$ is the uniform distribution and $k = \frac{1}{in\_features\_l}$, $in\_features\_l$ being the size of the input of layer $l$. For 10 000 steps, actions are sampled uniformly from [-$F_{max}$, $F_{max}$], $F_{max}$ being equal to 1 N, without concerting the policy or Q-functions. Once this fully exploratory phase is completed, the agent starts using ANNs. At each time step $t$, the agent observes the current state of the system $s_{t}$ (composed of the position, the velocity and the target state of each mass) and chooses a force $F_{e}(t)$ in the interval [-$F_{max}$, $F_{max}$] in consequence. The selected force, to which is added a noise taken from a Gaussian distribution of mean 0 and standard deviation 0.1, brings the system to a new state $s_{t+1}$ computed by Runge-Kutta method of order 4. At each time step (constant $F_{e}(t)$), the RK4 method is done through 10 successive iterations for a total duration of 0.1 s. Each of those iterations changes the state of the system. Once the numerical resolution is completed, the agent receives a reward $r_{t}$ given by function [\ref{eq:reward}].
\begin{align}
    r_{t} = -\sum_{i=1}^{N} (u_{i}^{(t+1)} - \delta_{i}^{(target)}) -\frac{1}{2}\sum_{i=1}^{N} \dot{x_{i}}^{(t+1)}
    \label{eq:reward}
\end{align}
where $target$ is a variable equal to 0 or 1, $u_{i}^{(t+1)}$ and $\dot{x_{i}}^{(t+1)}$ are respectively the displacement and velocity of the ith mass at time t+1. \\
The experience above is stocked in the replay buffer, which possesses a finite maximal size of 1e6 experiences. Each new experience overwrites the oldest stored one when the buffer is full. This process allows the continuous improvement of the available training dataset during training. The Q-functions and the policy are then updated. The Q-functions are updated at every step, while the policy is updated every two steps. Both the Q-functions and the policy are updated using the Adam algorithm~\cite{kingma2014adam} with a learning rate of 0.001 and a batch size of 100 experiences randomly sampled from the Replay Buffer. The operation goes on until either the agent reaches the goal, at which point it receives a reward of 50, or 200 steps are exceeded. At this stage, the environment is reset, giving place to a new episode. This algorithm is repeated for a predefined number of episodes.

\begin{figure}[h!]
    \centering
    \includegraphics[width=0.7\linewidth]{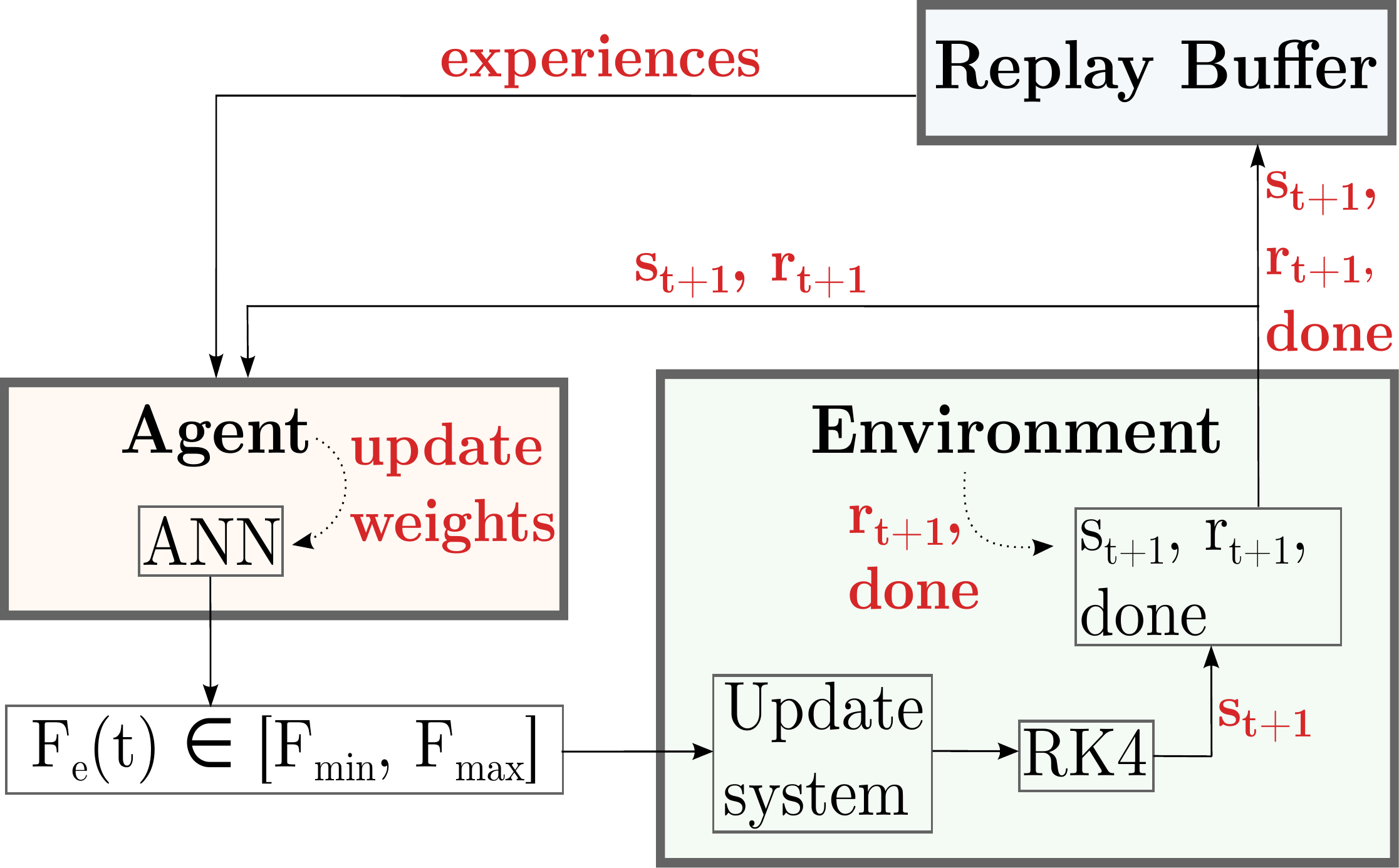}
    \caption{Architecture of the learning protocol.}
    \label{fig:learning_protocol}
\end{figure}

The fixed parameters of the environment, the hyper-parameters of the agent and the architecture of the neural networks of the policy and the Q-function are summarized in tables~\ref{table:env_param},~\ref{table:policy_q_func_models} and~\ref{table:hyperparameters}.\\

\begin{table}[]
\centering
\begin{tabular}{||c | c||} 
 \hline
 Parameter & Value \\ [0.5ex] 
 \hline\hline
 $F_{max}$ & 1 (N) \\ 
 $m$ & 1 (Kg) \\
 $k$ & 88.7 (N/m$^3$) \\ 
 $\delta^{(0)}$, $\delta^{(1)}$ & $\mathcal{O}$(1e-2 (m))\\
 dt & 0.1 (s) \\
 n$\_$res & 10 \\
 max$\_$episode$\_$len & 200 \\
 success$\_$pos & 0.005 (m) \\
 success$\_$vel & 0.01 (m/s) \\
 success$\_$r & 50 \\
 penalty$\_$pos & 1 \\
 penalty$\_$vel & $\frac{1}{2}$ \\[1ex]
 \hline
\end{tabular}
\caption{\label{table:env_param}Environment parameters. dt : discretization time, n$\_$res : number of iterations for the numerical resolution, max$\_$episode$\_$len : maximum number of steps per episodes, success$\_$pos : success condition on the position, success$\_$vel : success condition on the velocity, success$\_$r : success reward, penalty$\_$pos : penalty coefficient on the position, penalty$\_$vel : penalty coefficient on the velocity.}
\end{table}

\begin{table}[]
\centering
\begin{tabular}{||c | c||} 
 \hline
 Hyper-parameter & Value \\ [0.5ex] 
 \hline\hline
 Policy optimizer & Adam \\ 
 Policy learning rate & 0.001 \\
 Q-function optimizer & Adam \\ 
 Q-function learning rate & 0.001 \\
 $\gamma$ & 0.99 \\
 Replay buffer size & 1e6 \\
 Exploration time & 10 000 (steps)\\
 Batch size & 100 \\
 Policy update interval & 2 (steps)\\
 Q-function update interval & 1 (step)\\[1ex] 
 \hline
\end{tabular}
\caption{TD3 agent hyperparameters.}
\label{table:hyperparameters}
\end{table}

\begin{table}[]
\centering
\begin{tabular}{||c | c||} 
 \hline
 Policy layer & Policy activation \\ [0.5ex] 
 \hline\hline
 Linear (400) & Relu \\ 
 Linear (300) & Relu \\
 Linear (1) & Tanh \\ [1ex] 
 \hline\hline
 Q-function layer & Q-function activation \\ [0.5ex] 
 \hline\hline
 Linear (400) & Relu \\ 
 Linear (300) & Relu \\
 Linear (1) & None \\ [1ex] 
 \hline
\end{tabular}
\caption{Policy and Q-function models.}
\label{table:policy_q_func_models}
\end{table}

\section*{Four coupled bi-stable spring-mass units}

\begin{figure}[h!]
    \centering
    \includegraphics[width=0.9\linewidth]{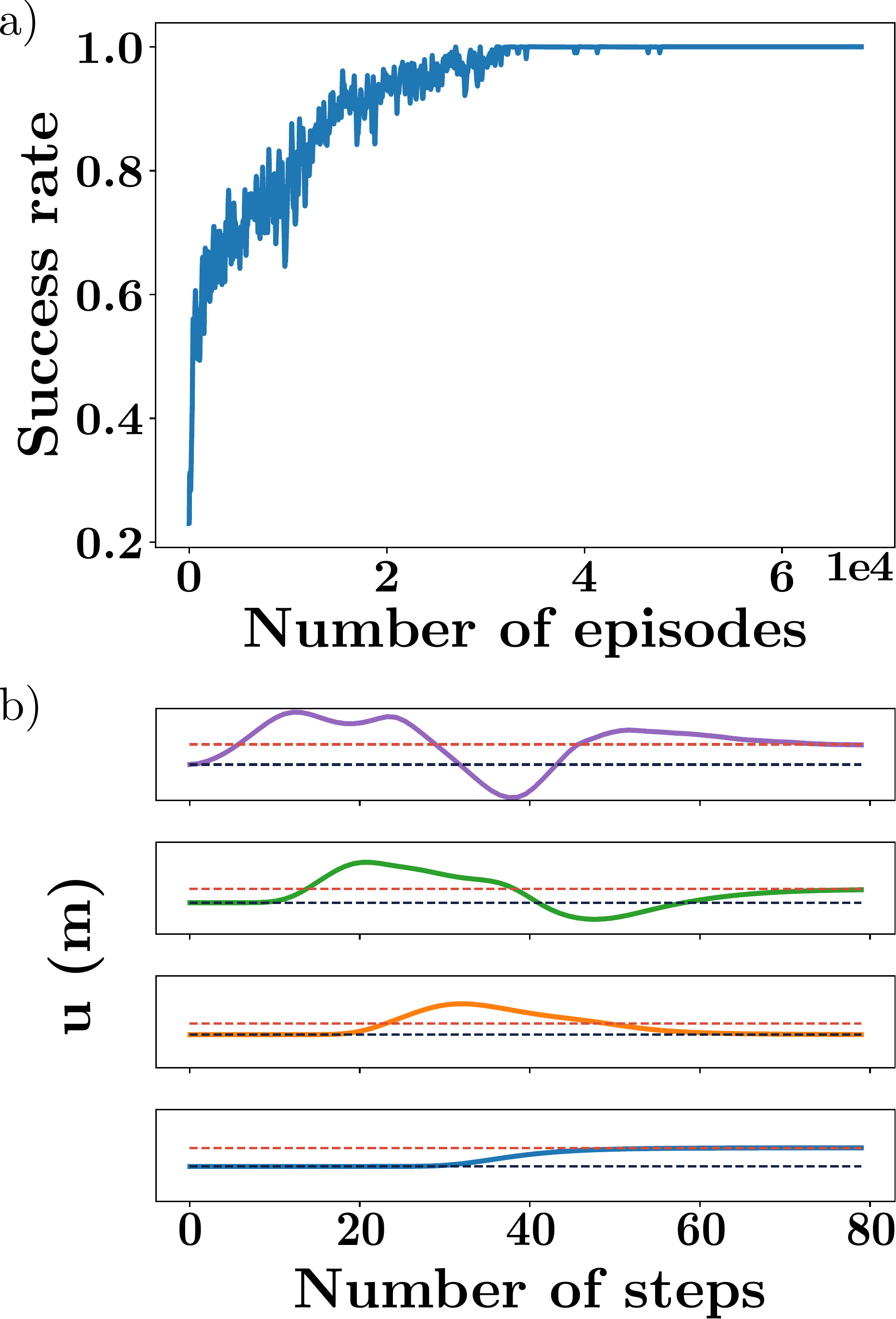}
    \caption{Learning dynamics of the RL agent on the model composed of four coupled bi-stable spring-mass units for $\eta$ = 2 Kg/s.
    a) Evolution of the success rate during training. The ANN were initialized randomly.
    b) Deformation of each bi-stable spring during the transition {\tt 0000} $\rightarrow$ {\tt 1011}. The colors blue, orange, green and purple correspond to the deformation of the first, second, third and fourth spring, respectively. The subplots for the deformations all have the same height with edge values [-0.14, 0.36] meter.  The dashed lines represent the stable equilibria $\delta^{(0)}$ (black) and $\delta^{(1)}$ (red).}
    \label{fig:four_masses}
\end{figure}

We trained a multistable chain composed of four bi-stable spring-mass units with the specific choice of disorder $\delta_1^{(0)}$ = 0.050, $\delta_1^{(1)}$ = 0.050, $\delta_2^{(0)}$ = 0.040, $\delta_2^{(1)}$ = 0.020, $\delta_3^{(0)}$ = 0.030, $\delta_3^{(1)}$ = 0.045, $\delta_4^{(0)}$ = 0.055, $\delta_4^{(1)}$ = 0.055 (m). 
This choice leads to four GoE states ${\tt 0010}$, ${\tt 1010}$, ${\tt 0011}$ and ${\tt 1011}$. 
The training was done with $\eta$ = 2 Kg/s.
All other physical parameters, hyperparameters, and training protocols were kept identical to the three springs training. 
The training dynamics and the deformation of the springs during the transition {\tt 0000} $\rightarrow$ {\tt 1011} proposed by the trained networks is shown in Fig.~\ref{fig:four_masses}.

\section*{Data Availability}

The code used to produce the results of this study is available online~\cite{RepoGit}.

\end{document}